\documentclass[sigconf,screen]{acmart}
\pdfoutput=1

\usepackage{amsmath}
\usepackage{subfigure}
\usepackage{epsfig}
\usepackage{float}
\usepackage{setspace}
\usepackage[ruled,vlined]{algorithm2e}
\usepackage{tabularx,booktabs}
\usepackage{graphicx}
\usepackage{makecell}
\usepackage{dcolumn}
\usepackage{array}
\usepackage{collcell}
\usepackage{comment}
\usepackage{url}
\usepackage[utf8]{inputenc}
\usepackage[normalem]{ulem}
\usepackage{changepage,threeparttable} % for wide tables
\useunder{\uline}{\ul}{}

\usepackage{xcolor}

\newcolumntype{H}{>{\setbox0=\hbox\bgroup}c<{\egroup}@{}}

\copyrightyear{2021} 
\acmYear{2021} 
\setcopyright{rightsretained} 
\acmConference[ESEC/FSE '21]{Proceedings of the 29th ACM Joint European Software Engineering Conference and Symposium on the Foundations of Software Engineering}{August 23--27, 2021}{Athens, Greece}
\acmBooktitle{Proceedings of the 29th ACM Joint European Software Engineering Conference and Symposium on the Foundations of Software Engineering (ESEC/FSE '21), August 23--27, 2021, Athens, Greece}\acmDOI{10.1145/3468264.3468563}
\acmISBN{978-1-4503-8562-6/21/08}

\begin{document}

\title{Sustainability Forecasting for Apache Incubator Projects}

\author{Likang Yin}
\affiliation{
\institution{DECAL and CS Department}
  \institution{University of California, Davis}}
\email{lkyin@ucdavis.edu}

\author{Zhuangzhi Chen}
\affiliation{
\institution{College of Information Engineering}
  \institution{Zhejiang University of Technology}}
\email{zhuangzhichen@foxmail.com}

\author{Qi Xuan}
\affiliation{
\institution{College of Information Engineering}
  \institution{Zhejiang University of Technology}}
\email{xuanqi@zjut.edu.cn}

\author{Vladimir Filkov}
\affiliation{
\institution{DECAL and CS Department}
  \institution{University of California, Davis}}
\email{vfilkov@ucdavis.edu}

\begin{comment}
\author{Likang~Yin*,
        Zhuangzhi~Chen,
        Qi~Xuan,
        Vladimir~Filkov*% <-this % stops a space
%\IEEEcompsocitemizethanks{\IEEEcompsocthanksitem L. Yin and V. Filkov are with DECAL and the Computer Science Department, University of California at Davis, Davis, CA 95616, USA.\protect\\
% \\ is fragile and will error, could use \hfil\break instead.\\
\footnote{lkyin@ucdavis.edu; vfilkov@ucdavis.edu}
%\IEEEcompsocthanksitem Zhuangzhi Chen and Qi Xuan are with College of Information Engineering, Zhejiang University of Technology, 288 Liuhe Road, Hangzhou 310023, China.}
}
\end{comment}


\begin{abstract}
Although OSS development is very popular, ultimately more than 80\% of OSS projects fail. Identifying the factors associated with OSS success can help in devising interventions when a project takes a downturn.
OSS success has been studied from a variety of angles, more recently in empirical studies of large numbers of diverse projects, using proxies for sustainability, e.g., internal metrics related to productivity and external ones, related to community popularity. The internal socio-technical structure of projects has also been shown important, especially their dynamics.
This points to another angle on evaluating software success, from the perspective of self-sustaining and self-governing communities.

To uncover the dynamics of how a project at a nascent development stage gradually evolves into a sustainable one, here we apply a socio-technical network modeling perspective to a dataset of Apache Software Foundation Incubator (ASFI), sustainability-labeled projects. To identify and validate the determinants of sustainability, we undertake a mix of quantitative and qualitative studies of ASFI projects' socio-technical network trajectories.
We develop interpretable models which can forecast a project becoming sustainable with 93+\% accuracy, within $8$ months of incubation start. 
Based on the interpretable models we describe a strategy for real-time monitoring and suggesting actions, which can be used by projects to correct their sustainability trajectories. %The data and scripts can be found: \url{https://doi.org/10.5281/zenodo.4564072}.
\end{abstract}

\maketitle

\begin{comment}
\section*{TODOs}
\begin{itemize}
\item Include distribution plot in the variable stats table.
\end{itemize}
\end{comment}

\section{Introduction}

Open source has democratized software development. Developers flock to OSS projects hoping to add certain functionality, contribute to a worthy goal, and sharpen their skills.
However, more than 80\% of OSS projects become abandoned over time, especially the smaller and younger projects~\cite{schweik2012internet}. 
Certainly not all OSS projects are meant to be widely used or even to persist beyond a college semester. 
However, even the large, popular OSS software, widely used in our daily lives and by fortune 500 companies, started out as small projects. 
Thus, from a societal perspective it is important to ask: Why do some nascent OSS projects succeed and become self-sustaining while others do not~\cite{raja2012defining}? And can the latter be helped?

To aid developer communities to build and maintain sustainable OSS projects, nonprofit organizations like the Apache Software Foundation (ASF) have been established. 
ASF runs the ASF Incubator (ASFI)~\cite{duenas2007apache}, where nascent projects aiming to be a part of the ASF community are provided with stewardship and mentor-like guidance to help them eventually become self-sustaining and even top-level projects in ASF. 
Projects in ASFI, called \emph{podlings}, are  required to adhere to ASF rules and regulations, including keeping all commits and emails public.
When certain conditions are satisfied, project developers and ASF committees decide if a podling should be \emph{graduated},  referred to as a `successful' sustainability outcome. Otherwise they get \emph{retired}. 
Per ASFI, \textit{A major criterion for graduation is to have developed an open and diverse meritocratic community. Graduation tests whether a project has learned enough and is responsible enough to sustain itself as such a community}\footnote{\url{https://incubator.apache.org/guides/graduation.html}}.

Podlings in ASFI receive a mentor, file monthly reports, and get feedback.  In spite of this support, many podlings fail. 
Most ASFI committers do not lack coding expertise, but graduating from the incubator requires more: it asks for effective teamwork and sustainable, community development.
These requirements are most challenging to meet.
From comments in ASFI, we see that developers are confused by the expectation of 'The Apache Way', especially initially.
E.g., from project \textit{Flex} on 05 Jan 2012,
'...I think there is a need to keep things as simple as possible for people who [are] already confused with whats going on with the move to Apache...'.
The frustrations sometimes persist beyond the initial period.
A comment from project \textit{Flex} on 06 Jan 2012, states  
'...many people are confused and lost as to what to do. Who is providing that direction for them?'
In large part, understanding how to achieve the graduation requirements seems to be the culprit, likely due to the abstract nature of those concepts. 
Another reason is that comparison to others is difficult.
From project \textit{Rave} on 28 June 2011:
"[sporadic adherence to requirements] makes it very confusing and difficult to compare against what 
other projects as some are doing too little while others are doing too much..." 

Thus, there is a need to connect the proverbial dots on how to get from the point of entry into ASFI to checking all the graduation requirement boxes. 
That brings us to the motivation of our paper. The extrinsically labeled ASFI dataset offers heretofore unavailable, fine-grained records of historical trajectories of projects under policies and regulations of ASFI.
We posit that:

\vspace{0.07in}
\fbox{ \parbox{3in}{The process that ASFI projects follow toward becoming sustainable can be modeled as a function of a small set of project features, so that the outcome (graduation/retirement) can be predicted early on, from the successes and failures of others, allowing for trajectory adjustment if needed.}}
\vspace{0.07in}

To deliver on that, in addition to the historical, outcome-labeled ASFI data, we also need a theoretical framework that can capture the complexity of OSS development.
Over the past 30 years research in organizational behavior and management has documented the evolution in project management practices~\cite{chaikalis2014forecasting}, as they have moved toward more successful models~\cite{andersen2009organizational}. 
The \emph{socio-technical} view of an organization~\cite{lin2006effects} has emerged as one of the more successful hybrid models recognizing the benefits that integrated treatment of the technical aspect (code, machines, device, etc.) and the social aspect (people, communication, well-being, etc.) has on an organization~\cite{palyart2017study}. 
Likewise, OSS projects have been effectively studied from the \emph{socio-technical system} (STS) perspective~\cite{amrit2010exploring}, with the social side capturing humans and their communication channels, and the technical capturing the content and structure of the software~\cite{sack2006methodological}.

\begin{figure}[tb]
    \centering
    \includegraphics[width= \linewidth]{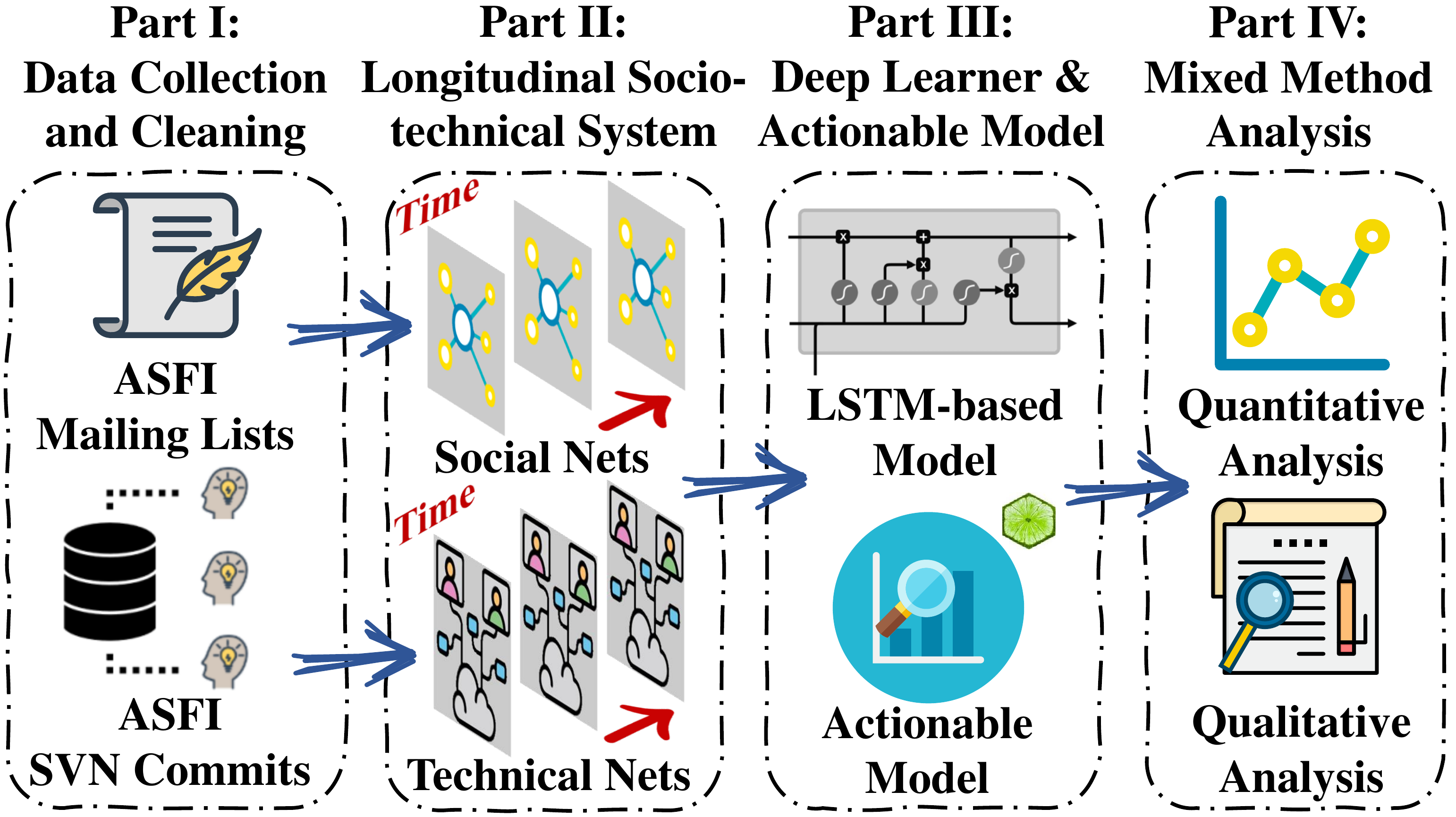}
    \caption{The workflow of our mixed-methods study.}
    \label{workflow}
\end{figure}

Here, inspired by the socio-technical systems modeling perspective, and the availability of extrinsically labeled historical data of project sustainability from ASFI, our goals are: (1) to identify socio-technical features distinguishing projects that graduate from the ASF Incubator from those that do not (i.e., find the determinants of OSS project sustainability), and (2) to build temporal forecasting models that can predict sustainability outcomes at any time point in the project development, and thus (3) to offer practical and timely advice on intervening to correct a project's course, especially early on.
To approach these goals, we conduct a mix of quantitative and qualitative empirical studies. 
We start by gathering project technical traces (commits) and social traces (emails) from the ASF Incubator website~\cite{ASFIP}. 
From those, we construct the temporal social and technical networks for each project, and perform exploratory data analyses, deep-dive case studies, build an accurate forecasting model, and finally implement the interpretable model presenting timely advice. We illustrate our workflow in~Figure \textcolor{red}{\ref{workflow}}. 
Our contributions in this paper are:
\begin{itemize}
\item We provide a novel longitudinal dataset of hundreds of OSS projects' development traces under ASF regulation, with extrinsically labeled project sustainability status.
\item We propose the first OSS project sustainability forecast measure modeled from tens of socio-technical network and project features. Our model shows excellent predictive performance ($\ge 93\%$ accuracy as early as 8 months into incubation).
\item We find that ASF incubator projects with \uline{fewer but more focused} committers and \uline{more but distributed} (participating in asynchronous discussions) communicators are more likely to gain momentum to self-sustainability.
\item We describe a strategy for real-time monitoring of  the sustainability forecast for any project, derived from an interpretable version of our DNN model.
\end{itemize}

This paper is a first step toward showing that end-results of project trajectories can be effectively forecast and possibly corrected upward, if needed. 
Our motivation goes beyond ASFI as many more nascent projects fail outside of ASFI, so self-monitoring and self-adjustment may be more pertinent to them.
%which is future work. 

%In the rest of the paper, we present related work and theories, followed by the methods, results and discussion, takeaways for practitioners, and the conclusion sections.

\section{Background and Theories}
\label{background}
We present background on the Apache Software Foundation (ASF) and OSS success, and then we introduce related theories through which we generate our Research Questions.

\subsubsection{Apache Software Foundation Incubator}
\textit{Community over code} is the tenet of the Apache Software Foundation (ASF) community~\cite{gharehyazie2015developer}. 
Their belief is that if they take good care of the community, good software will emerge from it.

However, conflicts are ubiquitous in OSS projects~\cite{kasi2014minimizing}, and not even ASF can escape them. To minimize conflicts, ASF requires projects to make all communication publicly available on the mailing lists, summarized popularly as \textit{if it did not happen in the mailing lists, it did not happen}~\cite{rigby2007can}. The communication records benefit developers as they reflect on previous decisions and trace their precursors, therefore improving efficiency and productivity.

The ASF community adopts a democratic way in many of their affairs. For example, contributors are invited to vote $+1$ (yes), $0$ (okay), or $-1$ (no) to project-wide changes. However, ASF committers can live in different time zones, and their response to a project decision can delay largely. Regarding that, ASF community adopts the \textit{Lazy Consensus}~\cite{marru2011apache}. 
%It states that if committers do not vote for more than certain time (typically 72 hours), it will be automatically counted as `+1' . 
Moreover, the ASF community also believes in \textit{Earned Rights}, that newcomers should be treated the same way if they have proven their technical skills.

The goal of the ASF Incubator (ASFI) is to help projects become self-sustained and eventually join ASF. Like many OSS developers, ASFI committers work at will and there are no formal obligations on them. Thus, ASFI projects are required to show they are able to recruit new committers, and fill existing technical debt~\cite{potdar2014exploratory}.
%That way, if any committers leave the project, it could still continue successfully. 
However, attracting new committers is difficult as they can be affected by both social and technical barriers~\cite{steinmacher2015social}. To address this issue, the ASFI community has established a set of specific rules that emphasize providing mentorship to newcomers~\cite{steinmacher2015systematic}.

During incubation, ASFI projects need to adopt ASF procedures to develop and cultivate the projects' community, and standardize their working style. 
To graduate from the incubator and finally become a part of the ASF, projects are required to demonstrate they can self-govern and be self-sustained~\cite{crowston2017core}. 
The specific requirement of sustainability can vary from one project to another~\cite{gamalielsson2014sustainability,duboc2019we}. 
%When project members decide their project can govern itself, they will initiate a free and public voting process on the project graduation proposal.
A project's graduation is ultimately approved for its self-sustainability by ASF's Project Management Committee via several rounds of public voting\footnote{\url{https://incubator.apache.org/guides/graduation.html}}. 
%When a project decides to be retired, the retirement does not necessarily mean that the existing code of the retired project cannot be re-used, it only indicates that there is no community responsible for maintaining that project.

\subsubsection{OSS Projects Success and Sustainability}
Recently, substantial work has focused on modeling the success of OSS projects~\cite{midha2012factors,piggott2013open,subramaniam2009determinants}. Even though there is no universally agreed definition of OSS success~\cite{gezici2019quality}, there are two main perspectives. 
The first one is from the development process viewpoint, which is often measured by technical metrics of software~\cite{ghapanchi2011taxonomy}, e.g., code defect density~\cite{rahmani2010study}, response time~\cite{mockus2002two}, and error resolution rate~\cite{kuan2001open}. The second one is more from the social angle, including contributor growth~\cite{zanetti2013rise}, community participation~\cite{mcdonald2013performance}, and communication patterns~\cite{wu2007investigating}. 
More specifically, K. Crowston et al.~\cite{crowston2006information} studied the operationalizing success measures under the context of FLOSS projects. 
N. Cerpa et al.~\cite{cerpa2010evaluating} provide survey evidence that factors in the logistic regression models are not as predictive as expected.
D. Surian et al.~\cite{surian2013predicting} identify discriminative rich patterns from socio-technical networks to predict project success in the context of \textit{SourceForge} projects. J. Coelho et al.~\cite{coelho2017modern} conduct mixed-method analysis on GitHub projects, and they find that most of modern OSS projects fail due to project characteristics (maintainability) and team issues (e.g., leaving of the main contributor).
M. Valiev et al.~\cite{valiev2018ecosystem} conducted studies on sustained activity in open-source projects within the PyPI ecosystem, and find that relative position in the dependency network has a significant impact on project sustainability.
None of these consider forecasting over time and thus are not useful for real-time monitoring, which is the major contribution of this work.
%\textbf{--- End ---}
%More specifically, researchers used technical and commercial success to measure the success of open source projects. E.g., Midha et al.~\cite{midha2012factors} find that language translations and user base are positively associated with OSS success. Param Vir. S notes that~\cite{singh2010small} some OSS community networks are characterized by small-world properties, allowing members in the community to access more quantity and variety of resources. Long. J~\cite{long2006understanding} shows that the role of core developers are much more crucial to both the community and OSS software development.

Although OSS success and sustainability measure similar aspects of projects~\cite{zanetti2012co}, they are, in fact, not the same thing. 
There are two main differences between the two. First, OSS success is measured statically while sustainability is measured dynamically. Second, sustainability is a measure related more to the human and social aspect (e.g., the ability to take responsible collective action, and an open and inclusive atmosphere, etc.) than the technical aspect (defect density, speed, and technical advantage, etc.).

Therefore, the sustainability of an OSS project becomes even more important when it is a part of a larger ecosystem~\cite{valiev2018ecosystem}. Such OSS projects are inter-dependent to each other, the sustainability and stability of one project can introduce tremendous network effect to its ecosystem.\footnote{A developer abruptly deleting the widely used, 11-line of \emph{left-pad} code, led to cascading disruption of other OSS projects in \textit{npm} ecosystem.} Therefore, the sustainability of OSS projects becomes even more significant as they can influence many other OSS projects in the ecosystem that rely on them.

\subsubsection{Socio-Technical Systems Theory}
Socio-technical structure plays an important role in achieving collective success in OSS projects~\cite{ducheneaut2005socialization,bird2006mining,wu2007investigating}.
%, therefore, we are also interested in studying how the socio-technical structure impact the sustainability of OSS projects. 
A Socio-Technical System (STS) comprises two entities~\cite{trist1981evolution}: the social system where members continuously create and share knowledge via various types of individual interactions, and the technical system where the members utilize the technical hardware to accomplish certain collective tasks. The theory of STS is often referenced when studying how the technical system is able to provide efficient and reliable individual interactions~\cite{herrmann2004modelling}, and how the social subsystem becomes contingent in the interactions and further affects the performance of the technical subsystem~\cite{fischer2011socio}.

OSS projects have been studied from the network view~\cite{ducheneaut2005socialization}. Gonz{\'a}lez-Barahona et al.~\cite{gonzalez2004community} proposed using technical networks, where nodes are the modules in the CVS repository and edges indicate two modules share common committers, to study the organization of ASF projects.

%As for defining the STS framework, Ropohl~\cite{ropohl1999philosophy} combine the views from both engineers and social scientists, and consider it as the intermediary entity that transfers the institutional influence to individuals. Fischer~\cite{fischer2012context} proposes a multidimensional approach for context-aware STS to ensure the precise delivery of information. More recently, Fischer~\cite{fischer2016exploring} proposes a theoretical framework for knowledge sharing in online collaborations, revealing that the public participation and decentralized contributions are the common features in a massive socio-technical system.

Moreover, in socio-technical systems, governance can be applied through long-term or short-term interventions. Smith et al.~\cite{smith2007moving} proposed two conceptual approaches: ‘Governance on the outside’ objectifies the socio-technical and is managerial in approach. ‘Governance on the inside’ is more reflexive about the role of governance in co-constituting the socio-technical. 
From that perspective, the ASFI community is a unique system that has both outside influence (regulations from ASF committee) and inside governance (motivated by the project managers).

%As for the practical usages of STS theory, Sutcliffe~\cite{sutcliffe2000requirements} proposed a theoretical STS model to analyze operational event flows in human-computer interactions. Yu. E~\cite{yu2011modelling} finds that when designing the technical support tools (e.g., SDEs) for developers, the human organizational context significantly impacts the actual benefits from the software. Guided by STS theory, Mubaroq et al.~\cite{mubaroq2019proactive} design online language learning forums, effectively relieved the unemployment problem in West Java. Another empirical evidence comes from Herrmann et al.~\cite{herrmann2016socio}, where they design an effective delivery service specialized for elderly people using framework from STS theory.

\subsubsection{Contingency Theory}

Contingency theory is the notion that there is no one best way to govern an organization. Instead, each decision in an organization must depend on its internal structure, contingent upon the external context (e.g., stakeholder~\cite{turner2004communication}, risk~\cite{cooke2002real}, schedule \cite{wearne1989study}, etc.). Joslin et al.~\cite{joslin2016impact} find that project success associates with the methodologies (e.g., process, tools, methods, etc.) adopted by the project. And not a single organizational structure is equally effective in all cases. As the organizational context changes over time, to maintain consistency, the project must adapt to its context accordingly. Otherwise, conflicts and inefficiency occur~\cite{barclay1991interdepartmental}, i.e., \textit{one size does not fit all}.

To address the conflicts caused by incompatible fitting to the project's context, previous work suggests thinking holistically. Lehtonen et al.~\cite{lehtonen2006three} consider the project environment as all measurable spatio-temporal factors when a project is initiated, processed, adjusted, and finally terminated, suggesting that the same factor can have an opposite influence on the projects under a different context. In the domain of software engineering, Joslin et al.~\cite{joslin2016impact} considers project governance to be part of the project context, concluding that project governance can impact the use and effectiveness of project methodologies.

%To measure how well a project fits into its context, Drazin et al.~\cite{drazin1985alternative} proposed a hierarchically conceptual framework ranging from low complexity to high complexity: selection, interaction, and the system method. M$\ddot{u}$ller~\cite{joslin2016impact} further summarizes the selection method as a qualitative measure on the relationship between context and performance, while the interaction method works more quantitatively. And the system method addresses the contextual factors (including both temporal and structural variables) holistically.

In the context of OSS projects seen as socio-technical systems, contingency theory implies that observing and tracking multiple facets/features of the projects may lead to more effective models of system evolution.

\section{Hypotheses and Research Questions}
\label{RQs}

Our goal is to build effective models for forecasting ASFI project sustainability. Here, we generate our hypotheses and formulate research questions based on prior work and the pertinent theories.

\subsubsection{Hypotheses}

STS theory suggests that publicly observable participation and decentralized contributions to software projects foster sustainable collaborations. 
The ASFI ecosystem is in a form of a typical STS where the technical activities build a shared code artifact and the social ones mediate knowledge and organizational details to individuals. 

Since all activities are logged, various socio-technical metrics can be calculated.
Our main hypothesis is that the STS formalism and the full availability of the projects' longitudinal digital traces, will make it possible to build an accurate model of sustainability.

According to contingency theory, no single organizational structure is equally effective under all circumstances. Thus, across ASFI projects, we expect to see that the same socio-technical factors may have different contributions to sustainability. 

Finally, we posit, per contingency theory, that the roles of some social-technical factors may vary over time. Moreover, we expect to see that similar ASFI projects can end with divergent outcomes (graduation or retirement) over time based on actions they've undertaken.

\subsubsection{Research Questions}
We formalize the above into our RQs, as follows. The first is a validation of contingency theory hypotheses, that there exist  measurable differences between graduated projects and retired projects, along with multiple features. Namely,

$\textbf{RQ}_{\textcolor{red}{1}}$ Are there significant differences among STS measures, between graduated projects and retired projects?

Next, STS theory holds that project sustainability is associated with social and technical network features. 
Contingency theory implies there will be multiple such features in play. Thus, we expect that a quantitative temporal model may be fitted well to the available ASFI data, so long as a sufficient number of features and projects are available. 

$\textbf{RQ}_{\textcolor{red}{2}}$ How well can we predict the sustainability based on temporal traces of ASFI projects? And, can we identify the determinants along with their weights and directions?

To make the model useful in practice it needs more than just accurate predictions of outcomes, it also needs to generate timely advice on whether the project should stay the course or implement specific corrective action to improve the graduation trajectory. We formulate this as:

%\textbf{***RQ3 and RQ4 may need editing***}

$\textbf{RQ}_{\textcolor{red}{3}}$  Can we monitor project sustainability status in a continuous manner? When and how should projects react to the monitoring advice?

%In the last research thrust, we are interested in the marginal projects which almost succeed but eventually failed or almost fail but finally land in graduation. We believe if we can predict or forecast such turning actions, we can aggregate them into \textit{just-in-time} advice/suggestion for ASFI project managers. We ask:

%\textbf{***RQ4 can relate to Quasi-experiment, but how can we reason it?***}

%\textbf{RQ4} What makes the turning points to the marginal projects? Can the monitoring tool capture such turning points, and thereby intervening the project development and change its outcome? 

%Lastly, as a project moves out of the ASFI, it opens up to a much wider potential contributor base. This transition can be substantial and further affect the project on its sustainability path. Thus, it is interesting to continue monitoring the projects even after they leave the incubator. Namely,

%\textbf{RQ4:} During the transition from a ASFI to ASF proper, are the projects experiencing notable changes in the contribution and communication rates? Can ASFI projects keep self-sustainable without ASFI community? 

\section{Data and Methods}

We collected historical trace data of commits, emails, incubation length, sponsor information, and incubation outcome for 263 ASFI projects, which have available archives of both commits and emails from 03/29/2003 to 10/31/2019. Among them, 176 projects have already graduated, 46 have retired, and 41 are still in incubation. The latter,  projects still in incubation, were not studied in this paper. 
%Regulations of Apache community make the ASFI data more trustworthy. Another benefit of using ASFI projects is that they have a extrinsic evaluated label indicating the incubation outcome given by experts from ASFI committee.
%Moreover, ASF committers often only focus very a few projects (and usually contribute to them for quite a long time) which they are really familiar with. This reduces the variance in project activity and makes productivity measures more trustworthy. Previous study suggests that the quality deficiencies in dataset can potentially impact the results of studies in empirical software engineering~\cite{bachmann2010process,cosentino2016findings}, therefore, when we study success of OSS projects, the presented ASF incubators dataset is a reliable to work with.

We collected the ASFI data from two sources: ASF mailing lists and SVN commits. The mailing list archives are open access and can be accessed through the archive web page, \url{http://mail-archives.apache.org/mod_mbox/}. They contain all emails and commits from the project's ASF entry date, and are current. 
We constructed URLs for individual project files in the ASF incubator as \href{http://mail-archives.apache.org/mod_mbox/}{\textit{Project URL}}. 
The project URLs use the  pattern: \texttt{project name/(YYYYMM).mbox}. 
For example, for project \emph{hama}, the full URL is \url{http://mail-archives.apache.org/mod\_mbox/hama-dev/201904.mbox}. 
Each such file contains a month of mailing list messages from the project, for the date specified in the URL. Here {\it dev} stands for `emails among developers'. 

We also collected other common mailing lists like `commits', `issues', `notifications', `users' (emails between users and developers or other users). However, we find that many projects, especially these over ten years old, which used SVN, used a bot in the `dev' mailing list to record all commits, thus a message from `dev' is not always an email. Similar emails were sent to the `commits' mailing list, which, thus, contains some emails. Therefore, we collected both `dev' and `commits' mailing lists files for the 222 graduated or retired ASF Incubator projects through the archive web page\footnote{Data and scripts are available: \url{https://doi.org/10.5281/zenodo.4564072}}. 

\subsubsection{Data Pre-processing}

ASF manages and records the communications among people by globally assigning an exclusive email name to each developer at the project-level. 
However, some developers still prefer to use their personal email/name instead of the assigned one, which in turn complicates the identification of distinct developers~\cite{zhu2019empirical}. We performed de-aliasing for those developers with multiple aliases and/or email addresses, as follows.
%, the merging for both username and email addresses are required. 
We first remove titles (e.g., jr.) and common words in the name (e.g., admin, lists, group) from usernames, then we match with both the original order and switched first/last name order whenever names contain exactly one comma to eliminate ambiguous styles. 
Then we match each developer with her/his multiple email addresses (if any). 
%For example, if two similar emails (often start with the same username but end with different domains) reply to the same discussion thread, it is likely that the two emails belong to the same developer.

Many projects, especially those over ten years old that used SVN, utilized a bot for extensive mailings (empirical evidence shows 26\% of popular GitHub projects use bots)~\cite{wessel2018power}, thus forming outliers in the dataset. Some broadcast emails are automatically generated by the issue tracking tool (e.g., JIRA), and no developer would reply to them. 
We eliminated the broadcast messages that no one replied to, and we find many of them are generated by automated tools. 
We find some developers contributed many commits by directly changing/uploading massive non-source code files (e.g., data, configuration, and image files).
Since those can form outliers in the dataset, commits to files with extension data: .json, .xml, .yml, .yaml, .jar; text/configurations: .config, .info, .ini, .txt, .md, and image: .jpg, .gif, .pdf, .png are eliminated.

As result, we identify 21,328 unique contributors (who either committed code or posted emails). Among them, 1,469 only committed code, and 18,205 only posted/replied to discussions without committing code. The remaining 1,654 contributors engaged in both activities. 
We identify 2,764,309 commits, modifying a total of 404,455 source code files. 
We collect 879,812 emails, from them we identify 19,859 developers who participated in discussions (by sending or receiving emails). 
Among them, 19,573 proactively engaged in discussion activities (i.e., sending emails), the remaining 286 developers collaborated in a passive way (only received emails).

\subsubsection{STS and Socio-technical Networks}
We use socio-technical networks to anchor our study of OSS STS.  
Network science approaches have been prominent in studying complex dynamics of OSS projects~\cite{bird2009putting,surian2013predicting}, although the specific definition may vary with domain context~\cite{meneely2011socio}.

In this paper, we define the projects' socio-technical structure using social (email-based) and technical (code-based) networks, induced from their emails to the mailing lists and commits to source files, as follows.
Similar to the approach by Bird et al.~\cite{bird2006mining}, we form a social (email) network for each project, at each month, from the communications between developers: a directed edge from developer \textit{A} to \textit{B} exists if \textit{B} has replied to \textit{A}'s post in a thread or if \textit{A} has emailed \textit{B} directly (which is contained in the ``in-reply-to'' field).
The technical (code) collaboration networks are formed for each project, at each month, by including an undirected edge between developer \textit{A} and developer \textit{B} if both developer \textit{A} and \textit{B} has committed to the same coding source file(s) \textit{F} that month (excluding the SVN branch names).
%Thus, they are bipartite graphs of developers, on one side and the files they touch, on the other.

\subsubsection{Features/Metrics of Interest}
\label{variables}
The socio-technical and project features/variables that we chose for this study have been identified based on our discussion and consideration of the underlying theories. All our data is longitudinal. All metrics are aggregated over monthly intervals, for each project, from the start to the end of its incubation~\cite{zhang2016use}.
We started with 29 variables, given their statistics as Supplementary Material.

\uline{Variable Selection}
We used Lasso regression~\cite{tibshirani1996regression} (L1 regularization) to identify a smaller set of 18 linearly independent variables, plus the outcome, described in the following. 
We used $R$'s library~\textit{glmnet}~\cite{friedman2009glmnet} for the Lasso regression, with $\lambda=0.001$.

\uline{Outcome}: Graduation Status. Graduation \texttt{grad\_status} is a binary variable (0=`Retired' or 1=`Graduated') indicating the projects graduation status in the incubator, as discussed above. 

\uline{Longitudinal Project Metrics}:
The number of Active Developers \texttt{num\_act\_devs} is the count of contributors who have been active by either making commits or participating in discussions. 
Number of commits \texttt{num\_commits} is the count of source code commits made by all committers in the project. 
The process of excluding the commits that do not contain source code is described in the Data Section. 
The number of Emails \texttt{num\_emails} is the number of emails (including both thread starter emails and reply-to emails).
\texttt{num\_files} is the total number of unique source code files created during the incubation.
To measure the continuity of activities, we define \texttt{c\_interruption} and \texttt{e} as the sum of the time intervals of the top $3$ longest interruptions between successive commits and successive emails, respectively. 
%We also use monthly skewness ($skew\_c$, $skew\_e$) to measure if the corresponding activity (commits or emails) to measure activity asymmetry during the month. 
%They are calculated by taking non-parametric skew ((mean - median) / standard deviation) on the distribution of activity dates (i.e., the gap days between the beginning of the months and the current activity date). 
%A positive skew means the activities are front-loaded in the month, and a positive one the opposite.
The commit percentage \texttt{top\_c\_fract}  and email percentage  \texttt{top\_e\_fract}  are the percentages of respective activities performed by the top $10\%$ contributors.

\uline{Longitudinal Socio-Technical Project Metrics}:  
For each project network, for each month, we constructed the technical and social networks, and from them calculate the number of active nodes, \texttt{c\_nodes}, and edges \texttt{c\_edges} in the technical network; \texttt{e\_nodes} and \texttt{e\_edges} in the social network. 
The prefix \texttt{c\_} in a variable's name indicates it is of the technical (code) network, while the prefix \texttt{e\_} in a variable's name indicates it is of the social (email) network. 
Additionally, we calculated the mean degrees \texttt{c\_mean\_degree} and \texttt{e\_mean\_degree} (sum of all nodes' degree divided by the number of nodes) in the technical network and social network, respectively. 
We calculate the clustering coefficients \texttt{c\_c\_coef}, \texttt{e\_c\_coef} as the number of connected triplets divided by the number of all triplets in the corresponding monthly network. 
The long-tail-edness \texttt{c\_long\_tail}, \texttt{e\_long\_tail} is calculated as the degree of the $75th$ percentile of nodes in the network, for the monthly networks, in the technical and social network, respectively. 
To get a sense of the range and variability in these variables, we show them aggregated over all months and projects in Table~\textcolor{red}{\ref{stats}}.
%\footnote{Mins of zero are mostly due to the early retirement of one project (Kabuki)}.

\begin{table}[b] \centering 
\caption{Statistics of the 176 graduated and 46 retired projects in the ASFI dataset.  $c\_$ and $e\_$ correspond to technical networks and social networks, respectively.}
\label{stats}
\scalebox{0.91}{
\begin{tabular}{@{\extracolsep{5pt}}lHcccHHc} 
\\[-3.8ex]\hline 
\hline \\[-1.8ex] 
Statistic &  & \multicolumn{1}{c}{Mean} & \multicolumn{1}{c}{St. Dev.} & \multicolumn{1}{c}{5\%} &  &  & \multicolumn{1}{c}{95\%} \\ 
\hline \\[-1.8ex] 
\texttt{grad\_status} & 222 & 0.79 & 0.41 & 0 & 1 & 1 & 1 \\
%incubation\_month & 222 & 22.94 & 15.23 & 6.88 & 12.08 & 30.19 & 52.96 \\ 
\hline
\texttt{num\_files} & 222 & 1,821.87 & 3,346.23 & 122.45 & 517.2 & 2,068 & 5436.6 \\ 
\texttt{num\_emails} & 222 & 3,963.12 & 4,930.54 & 262.65 & 1,116 & 4,874.2 & 12463.6 \\ 
\texttt{num\_commits} & 222 & 12,451.84 & 27,373.41 & 453.8 & 2,321 & 13,562.8 & 36359.7 \\ 
\texttt{num\_act\_devs} & 222 & 121.23 & 119.85 & 25 & 50 & 146.5 & 415.05 \\ 
\hline
\texttt{c\_interruption} & 222 & 0.20 & 0.20 & 0.03 & 0.06 & 0.28 & 0.60 \\ 
\texttt{e\_interruption} & 222 & 0.11 & 0.14 & 0.01 & 0.04 & 0.14 & 0.32 \\ 
%skew\_c & 222 & 0.32 & 0.97 & $-$2.41 & $-$0.23 & 0.73 & 5.00 \\ 
%skew\_e & 222 & 0.03 & 0.98 & $-$10.58 & $-$0.32 & 0.38 & 2.43 \\ 
\hline
\texttt{top\_c\_fract} & 222 & 0.65 & 0.19 & 0.38 & 0.50 & 0.80 & 0.94 \\ 
\texttt{top\_e\_fract} & 222 & 0.71 & 0.11 & 0.49 & 0.66 & 0.79 & 0.85 \\ 
\hline
\texttt{c\_nodes} & 222 & 15.44 & 17.09 & 2 & 6 & 18 & 49.9 \\ 
\texttt{c\_edges} & 222 & 120.15 & 276.19 & 1 & 10 & 91 & 531.5 \\ 
\texttt{c\_c\_coef} & 222 & 0.78 & 0.25 & 0 & 0.8 & 0.9 & 1 \\ 
\texttt{c\_long\_tail} & 222 & 10.65 & 11.93 & 0 & 3 & 14 & 33.85 \\ 
%c\_triangles & 222 & 861.31 & 3,405.90 & 0 & 7 & 242.5 & 36,798 \\ 
\texttt{c\_mean\_degree} & 222 & 8.14 & 7.45 & 1 & 3.5 & 10.3 & 23.75 \\ 
\hline
\texttt{e\_nodes} & 222 & 113.38 & 115.07 & 22 & 45.2 & 130.8 & 408.15 \\ 
\texttt{e\_edges} & 222 & 399.23 & 562.38 & 47.1 & 117 & 447 & 1315.9 \\ 
\texttt{e\_c\_coef} & 222 & 0.43 & 0.10 & 0.28 & 0.38 & 0.50 & 0.58 \\ 
\texttt{e\_long\_tail} & 222 & 10.33 & 7.15 & 3 & 5 & 13 & 24.95 \\ 
%e\_triangles & 222 & 829.33 & 2,360.87 & 0 & 98.2 & 575.8 & 23,177 \\ 
%e\_bidirected\_edges & 222 & 312.40 & 436.24 & 2.00 & 88.62 & 358.50 & 3,331.00 \\ 
\texttt{e\_mean\_degree} & 222 & 6.07 & 1.91 & 3.88 & 4.73 & 6.96 & 9.62 \\ 
\hline \\[-2.8ex] 
\end{tabular}}
\end{table}

\begin{figure*}[tbp]
\centering
\subfigure[Incubation Months ($p < .001$)]{
\label{incubation_length} 
\includegraphics[width=0.18\linewidth]{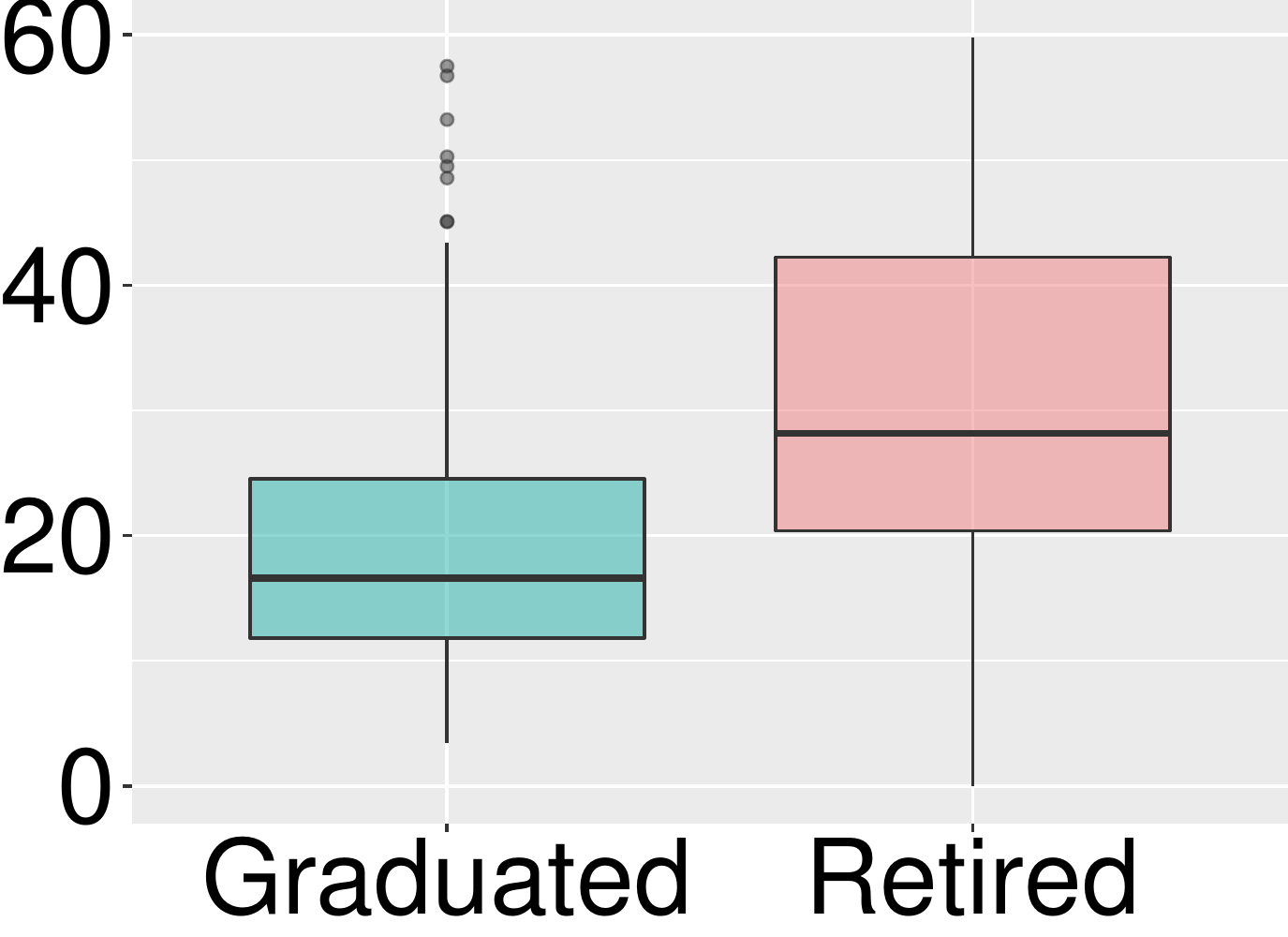}}
\subfigure[Num. of Commits ($p < .004$)]{
\label{num_commits}
\includegraphics[width=0.18\linewidth]{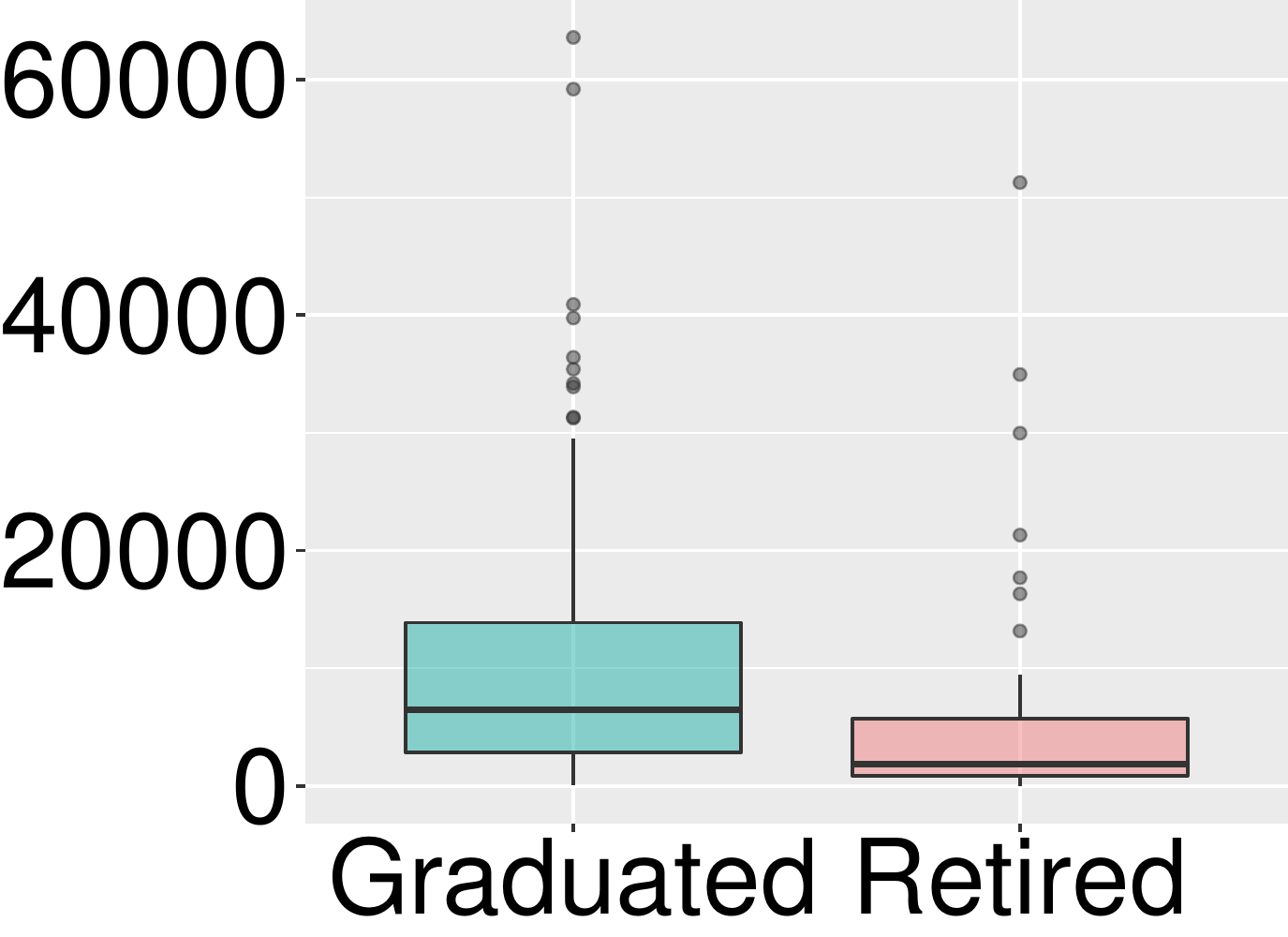}}
\subfigure[Num. of Emails ($p < .001$)]{
\label{num_emails} 
\includegraphics[width=0.18\linewidth]{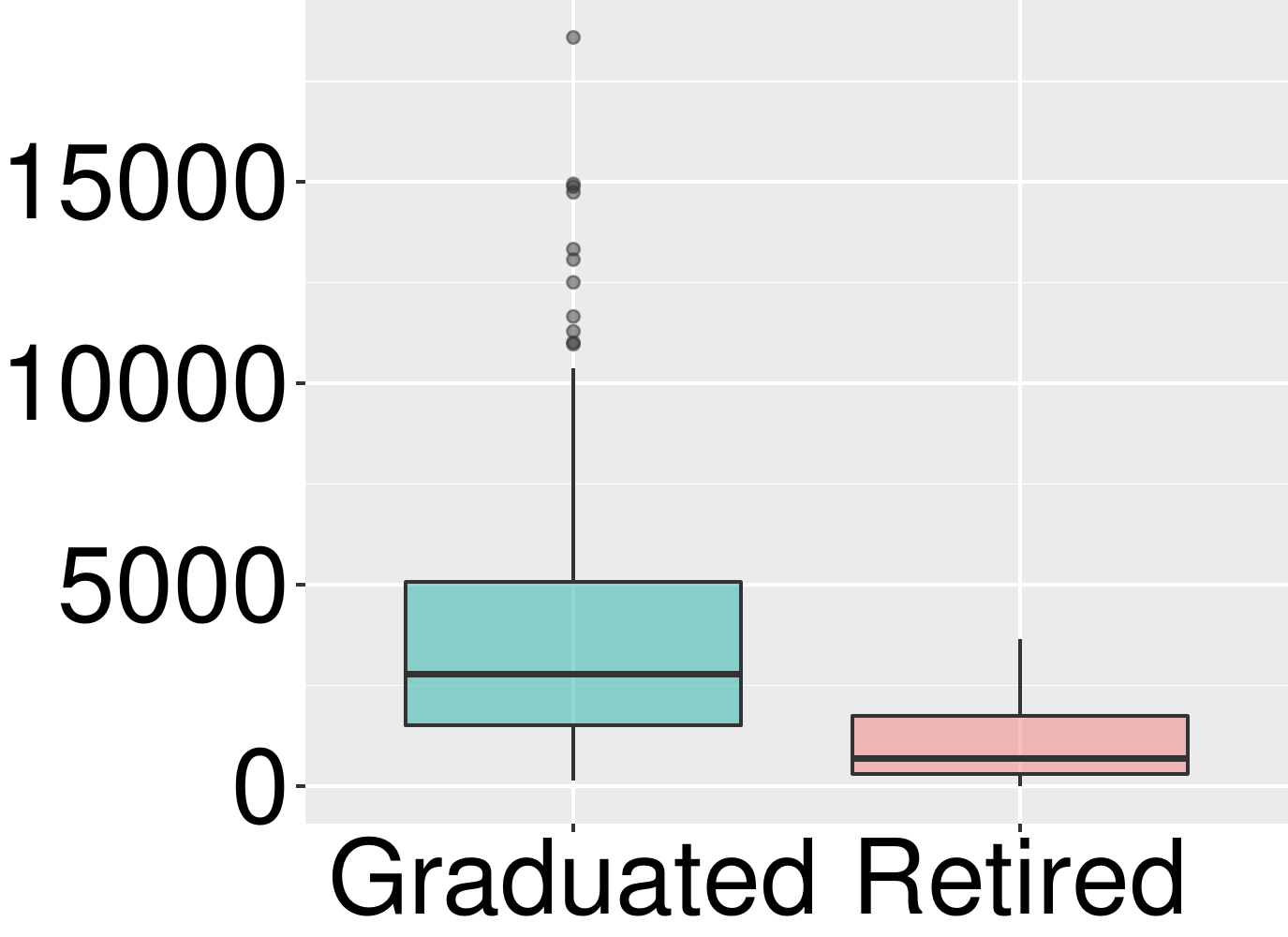}}
%\subfigure[Number of Files]{
%\label{num_files}
%\includegraphics[width=0.23\linewidth]{figs/num_files.pdf}}
\subfigure[TN Nodes ($p < .001$)]{
\label{c_nodes}
\includegraphics[width=0.18\linewidth]{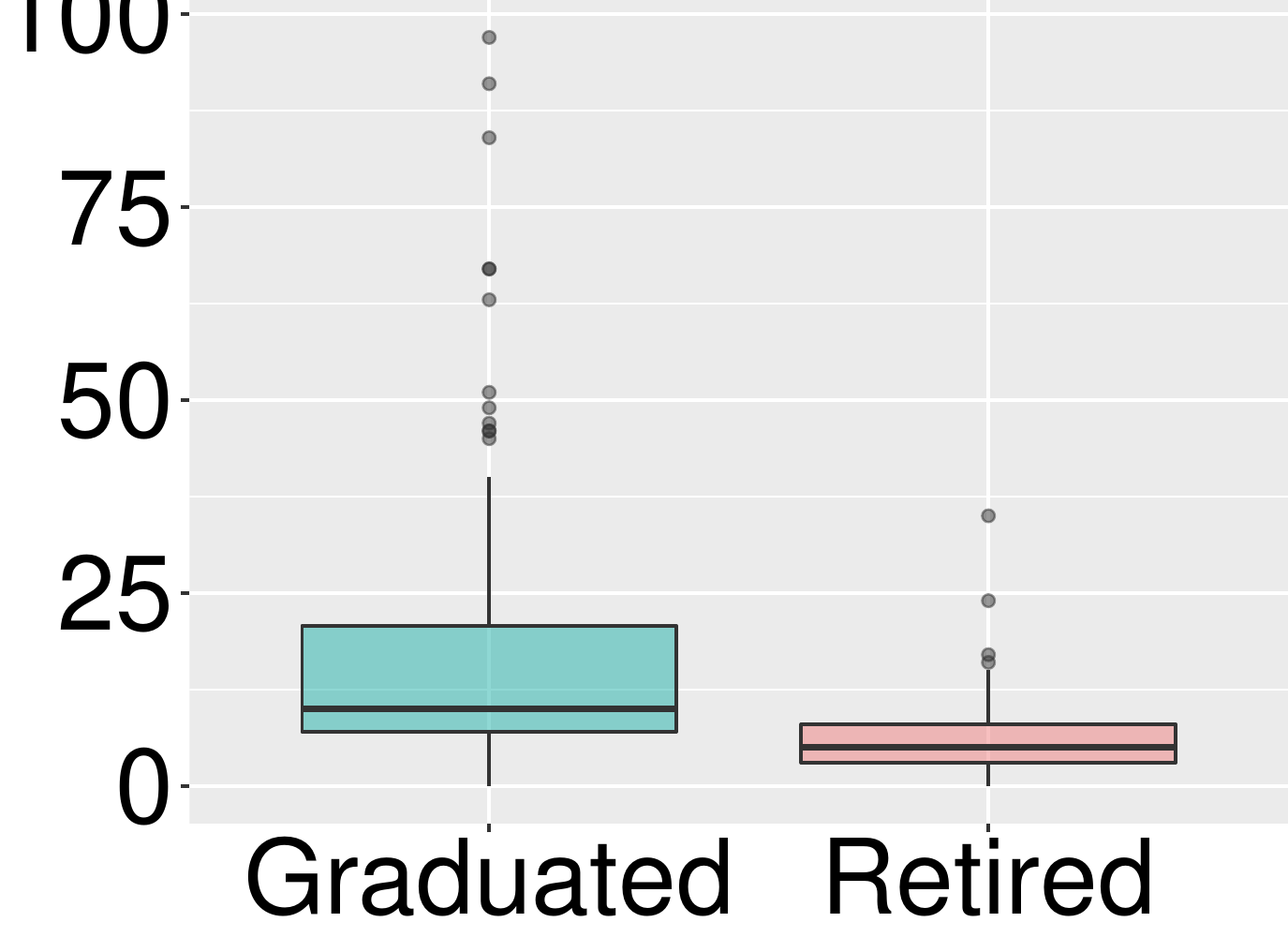}}
%\subfigure[Edges in Technical Network]{
%\label{c_edges}
%\includegraphics[width=0.23\linewidth]{figs/c_edges.pdf}}
\subfigure[SN Nodes ($p < .001$)]{
\label{e_nodes}
\includegraphics[width=0.18\linewidth]{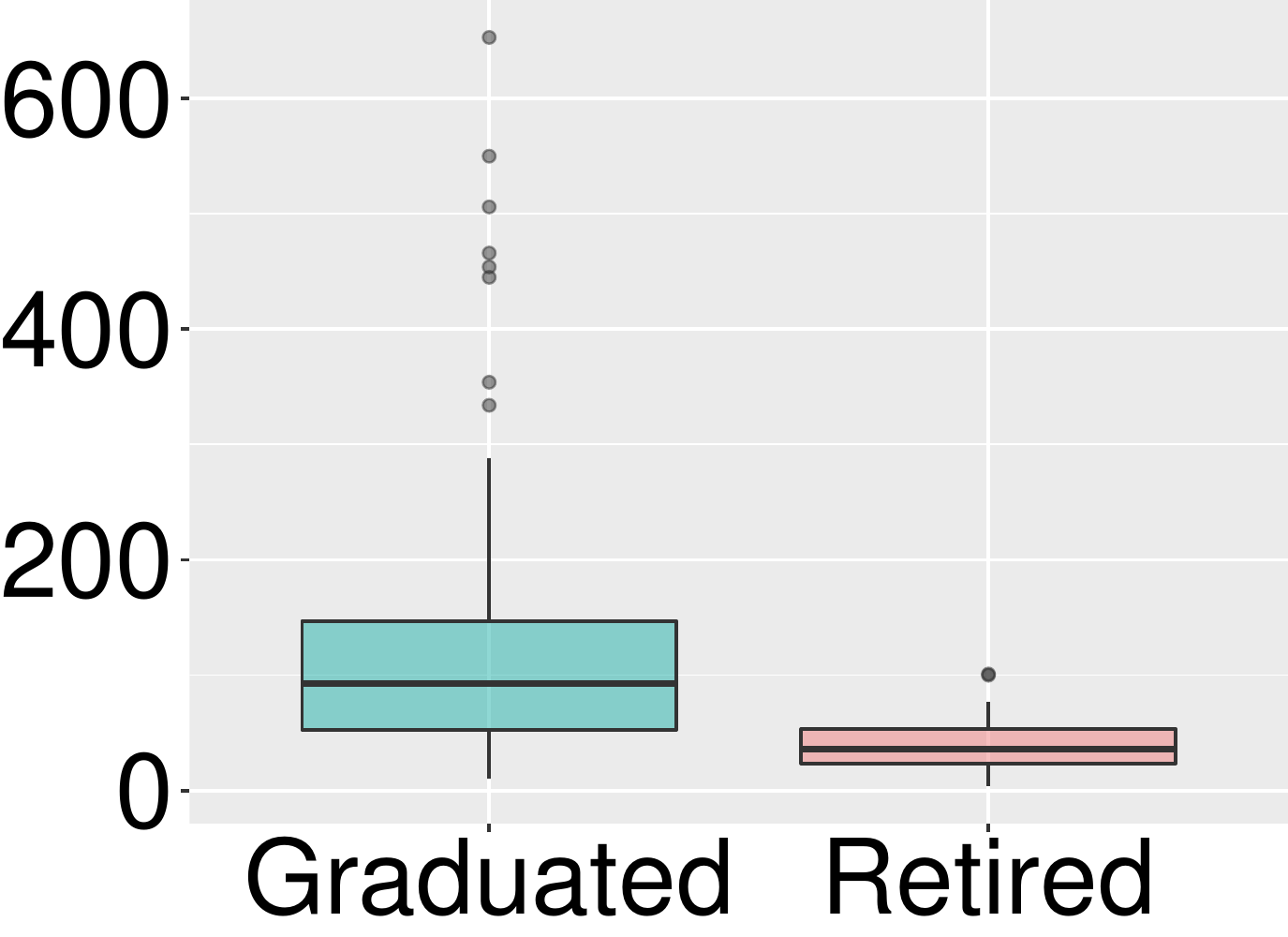}}
%\subfigure[Edges in Email Network]{
%\label{e_edges}
%\includegraphics[width=0.23\linewidth]{figs/e_edges.pdf}}
\caption{The descriptive variables between graduated projects (in green, left) and retired projects (in red, right). The corresponding p-value of the Student's t-test is in the brackets, suggesting significant statistical differences exist between them.}
\label{fig:stats}
\end{figure*}

\subsubsection{Modeling Approach} 
\label{NN}
%Compared with the general neural network, Recurrent Neural Network (RNN) is a neural network used to process sequential data. 
We needed a modeling approach able to learn and forecast from longitudinal data, have excellent performance, and be interpretable. 
Long Short-Term Memory (LSTMs) model is a variant of Recurrent Neural Networks (RNNs), designed to learn and model sequential data and is less sensitive to the problem of gradient disappearance and gradient explosion when training on long sequences. 

To obtain sequence data for each project, we aggregated historical ASFI records into monthly data, for each month from the incubation start date to the project graduation/retirement from the incubator. 
We interpret the monthly LSTM output probability as the \emph{graduation forecast}, i.e., the probability of the project eventually graduating.

%Our experimental setting is as follows: 
We prepared the data as follows. 
%Since in the ASFI dataset there are more graduated projects than retired projects (about 3 to 1 ratio), we first up-sample retired projects by a factor of 3 to get a balanced labeled dataset for training.
We randomly divided the projects into training and test sets in an 8-to-2 ratio. 
Because we have variables that are of very different magnitudes, and many of those are not normally distributed, thus we choose to use the \textit{MinMaxScaler} function to standardize all prediction variables.

We implemented a 3-layer LSTM model: a $64$ neurons LSTM layer, followed by a $0.3$ rate drop-out layer, and a dense layer with the \textit{softmax} function to yield the predicted outcome of the classification task (graduate/retire). 
During training, we used a binary cross-entropy as the loss function and $Adam$ as the optimizer. 
Since the length of the temporal data of each project varies, instead of using zero-padding, which could possibly introduce variance to the model, we chose to use a slower but more reliable way by only feeding one training sample at a time.
%We used 100 training epochs, and we visually confirmed that the loss converged around 50 epochs. 
We use the accuracy, precision, recall, and $F_1$-measure to evaluate the performance of the LSTM model using the \textit{classification\_report} function from the \textit{sklearn} package.

To get the \textit{graduation forecast} for a project at month $m+1$, we cap the project history at month $m$, i.e., we only use the first $m$ months in the model.
We interpret the outcome yield of the LSTM model as the \textit{graduation forecast}.
Projects are not being calculated and used in the prediction when the current time exceeds their incubation lengths.
We generated graduation forecasts for each month, and thus obtained the \textit{graduation forecast trajectories}. 
Repeating the above process 10 times, selecting different training/test split each time, produced our error bounds. 

\subsubsection{LIME-based Interpretable Model}
Black-box deep learning models, like LSTMs, are less ideal for decision making than interpretable approximations of deep learning models~\cite{molnar2020interpretable}.
One such approach is the Local Interpretable Model-agnostic Explanations (LIME) method~\cite{ribeiro2016should}. 
Given a pre-trained model and an input instance, LIME reasons how the black-box model yielded the output, by probing the model along each of the features.
LIME yields a magnitude and a sign (positive or negative) that characterize the contribution of each feature toward explaining the outcome.

\uline{Assumption} LIME assumes that any complex model is linear (i.e., interpretable) at a local scale. Therefore, given an input instance, LIME first artificially generates large enough samples that are presumably very close to the given sample (by some distance measure). 
Then, by training on the predictions of those newly generated samples given by the complex model, LIME can locally approximate the complex model using linear models, thereby presenting the coefficients of variables of the input instance.

\uline{Procedure}  
We first constructed a LIME explainer using the \textit{RecurrentTabularExplainer} function from the Python LIME package (\textit{version 0.2.0}).
That package was designed for explaining RNN-type models with tabular data. 
For each project, we use $explain\_instance$ function, setting the parameter \textit{num\_features} to the product of the incubation length and the number of features.
In this way, we can obtain the coefficients of all features over all time.
The parameter \textit{num\_samples} is set to 5,000 (by default), which is empirically sufficient for convergent results. 
Next, LIME probes the pre-trained LSTM model 5,000 times by feeding it the newly generated samples. 
LIME uses a similarity/distance function to measure the importance of each new sample on the locality of the instance to be explained. 
Lastly, LIME fits a weighted linear model dataset, and the explanations all come from the final linear model.
Since the LIME framework requires all samples to have the same shape, we divided our projects into several buckets where the projects have at least $n$ months of temporal data in the $n$-th bucket.

\uline{Project-specific vs Overall Modeling} We used the LIME results in two modeling ways: \emph{project-specific} level and \emph{overall} level.
In the former, we used LIME to obtain the monthly coefficient of each variable and then aggregated  them over all months to obtain a project-specific coefficient. 
In the latter, we aggregated project-specific coefficients over all ASFI projects to obtain the coefficients for each variable over all projects.

% how to process the source code files
\section{Results and Discussions}
%In this section, we apply the above methodologies to the research questions and discuss the corresponding results.
\subsection{$\textbf{RQ}_{\textcolor{red}{1}}$: Graduated vs. Retired Projects}
\label{RQ1}
To perform exploratory data analysis, we first contrast ASFI graduated and retired projects along the technical (code-based) and social (email-based) dimensions in our data.

Box-plot comparisons of the incubation length, number of commits, number of emails, nodes in the social networks, and nodes in the technical networks are shown in Figure~\textcolor{red}{\ref{fig:stats}}.
We observe notable differences as follows. 
The median incubation length (in months) of retired projects is significantly higher than of the graduated projects, suggesting that retirement is not an easy decision, and that perhaps necessary time is given to projects to change their trajectories and achieve graduation.

Graduated projects also tend to have more code commits and more email communications, implying that, in terms of criteria for graduation, the ASF community values both technical contribution (as commits) and social communication (as emails), and both of them may be of importance in building a sustainable community. 
Such results also motivate us to expand our research goals from descriptive data to inferential data with more complex network features. 

%Lastly, the number of emails tends to have higher variance than the number of commits, implying that the way how graduated/retired projects communicate differs in a larger part than the way how they work.
Across both the social and technical networks, the network size varies for the graduated projects, indicating that projects of any size can be sustainable, and it also suggests that project size can be used as a control in modeling.
The notable difference between graduated and retired projects, and the lack of variance in network sizes in the retired projects suggests recruitment difficulties in the latter, exposing them to significant risks as people leave.

\vspace{0.2cm}
\fbox{%
  \centering
  \parbox{0.92\linewidth}{
    \textbf{Answer to $\textbf{RQ}_{\textcolor{red}{1}}$:} We observe significant differences across multiple key measures between graduated projects and retired projects. Notably, retired projects tend to stay longer in the incubator than graduated ones, and the productivity and diversity of graduated projects are higher compared to retired projects. 
}%
}

\subsection{$\textbf{RQ}_{\textcolor{red}{2}}$: Interpretable Forecasting}
\label{RQ2}
\begin{figure}[t]
    \centering
    \includegraphics[width=\linewidth]{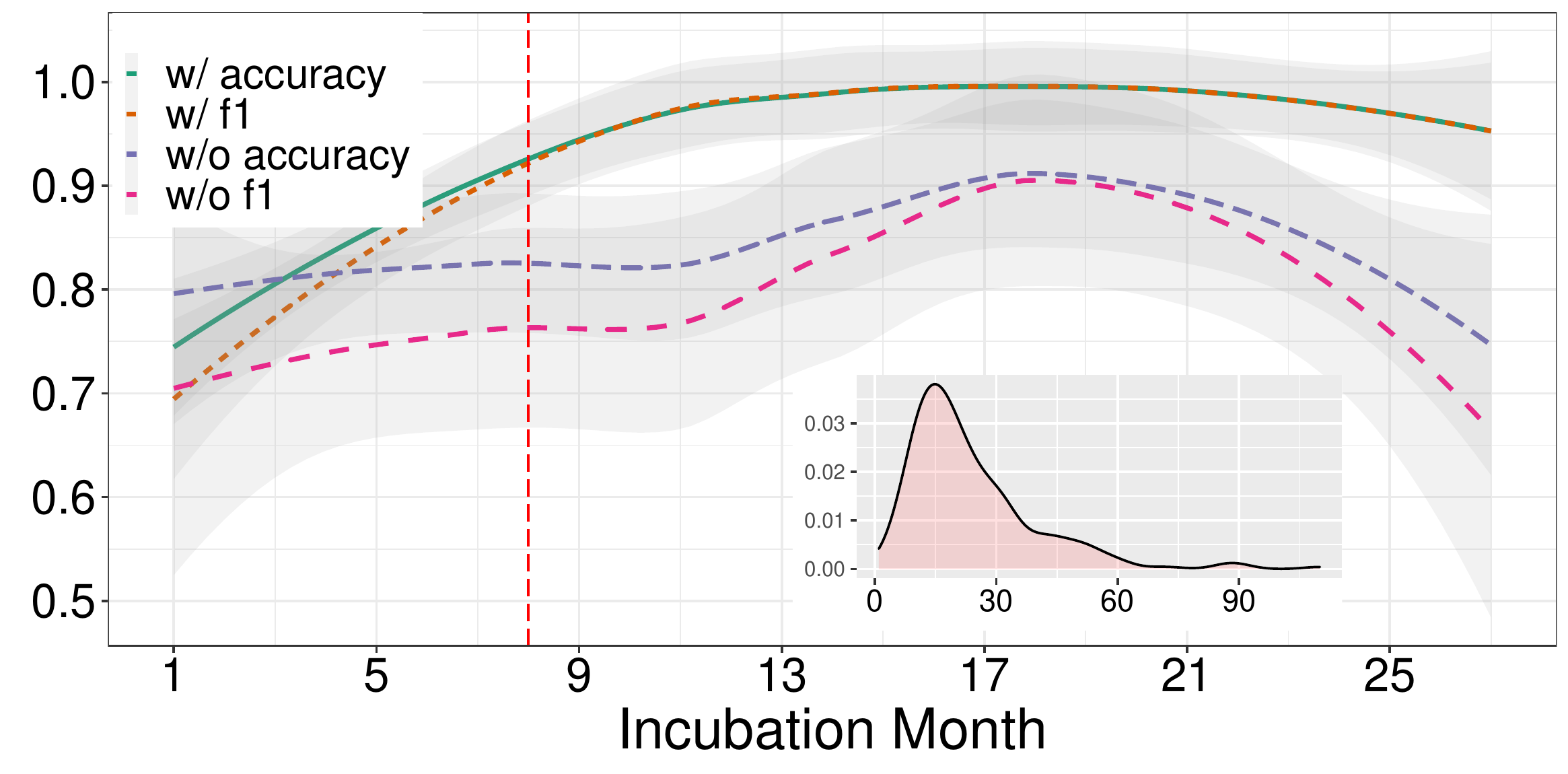}
    \caption{Performance metrics of the full LSTM model across incubation months (top 2 curves), showing the significant contribution of the socio-technical metrics. Curves are plotted using \textit{loess}. Grey area shows the standard errors. The red vertical line shows 93\% accuracy at 8 months of incubation. The inset shows project density over total incubation time.}
    \label{f1}
\end{figure}

Here we present the results of training an LSTM model on our ASFI data, and an LSTM-derived LIME model on the same data, using the methods as described above.
We use those models for forecasting by each month into the project the eventual graduation outcome. 
First, we show the performance curves of the LSTM model, over time, in Figure~\textcolor{red}{\ref{f1}}. 
The overall accuracy, and F1-value, and their standard errors, for the full model and the model without the socio-technical variables, suggest an excellent and stable predictive performance of the trained LSTM model, with significant contribution from the socio-technical networks.
As early as month 8 the accuracy is 93\%, and staying above after that. 

The total incubation time varies significantly across ASFI projects. 
While for most projects the incubation time is between $8$ months and $25$ months, some spend more than $30$ months in incubation, while others only 6 months, as shown in the inset plot in Figure~\textcolor{red}{\ref{f1}}. 
Thus, the model's performance decreases for projects with below $8$ and above $25$ incubation months, due to insufficient data above and below those values.

\begin{figure}[t]
    \centering
    \includegraphics[width= \linewidth]{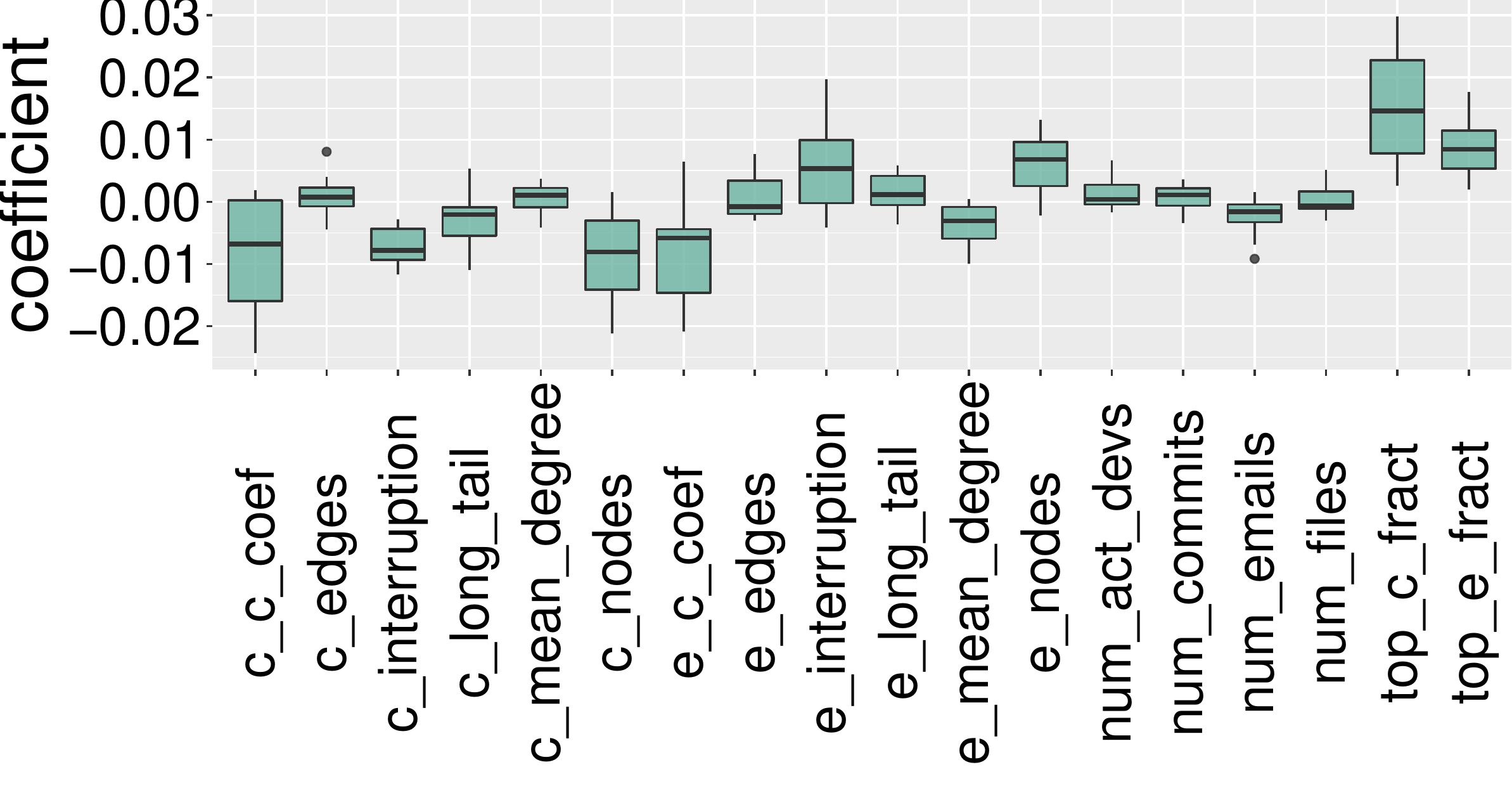}
    \caption{The coefficients of all variables from a graduated project (`Empire-DB'), aggregated  over all incubation months, showing that LIME delivers stable estimation at the project-level
    }
    \label{fig:lime_result2}
\end{figure}

Next, we apply LIME to understand and interpret the LSTM model and derive regression-like coefficients for the features in the socio-technical networks. 
To illustrate how to interpret the results at a project-specific level, we show an example of LIME's output for a graduated project, `Empire-DB', in Figure~\textcolor{red}{\ref{fig:lime_result2}}.
Note that the coefficients are aggregated over all incubation months.
Looking at the median over all months provides model coefficient stability and avoids emphasizing very small effects which nevertheless dominate in some months.
%\textbf{Likang added}: 
The magnitudes of the coefficients tend to be small because the model predicts probabilities within [0, 1], and there are tens of them.

There, we see that the technical network clustering coefficient \texttt{c\_c\_coef} is negatively associated with successful graduation. 
A high clustering coefficient in the technical network of people and source files indicates a high overlap of developers' activities on the same files.
One possible reason for the negative effect introduced by \texttt{c\_c\_coef} is that, work may not be well distributed among the team members. 
Another reason might be that the artifact is not well-conceived. 
Yet a third reason can be that the number of developers on the project is small and they must all ``tend to fires'' wherever they might be. 
Interestingly, the fraction of commits \texttt{top\_c\_fract} and emails ($top\_e\_fract$) by the top 10\% developers who often are the influencers in projects are positively associated with graduation, implying that the efforts of the top 10\% help sustain the project. 
Figure~\textcolor{red}{\ref{fig:lime_result2}} also shows the coefficients of the features in both size and direction across all incubation months, which provides further insight, and confidence in the methods utility.

Next, by only counting the signs of the project-level coefficients across all projects, we can identify whether there is an overall consistent direction in which that feature is contributing to the prediction.
E.g., if a feature is consistently negative to the outcome across all projects, i.e., that feature's coefficient is negative in most project models, then it has the same, overall negative effect.

\begin{figure}
    \centering
    \includegraphics[width= \linewidth]{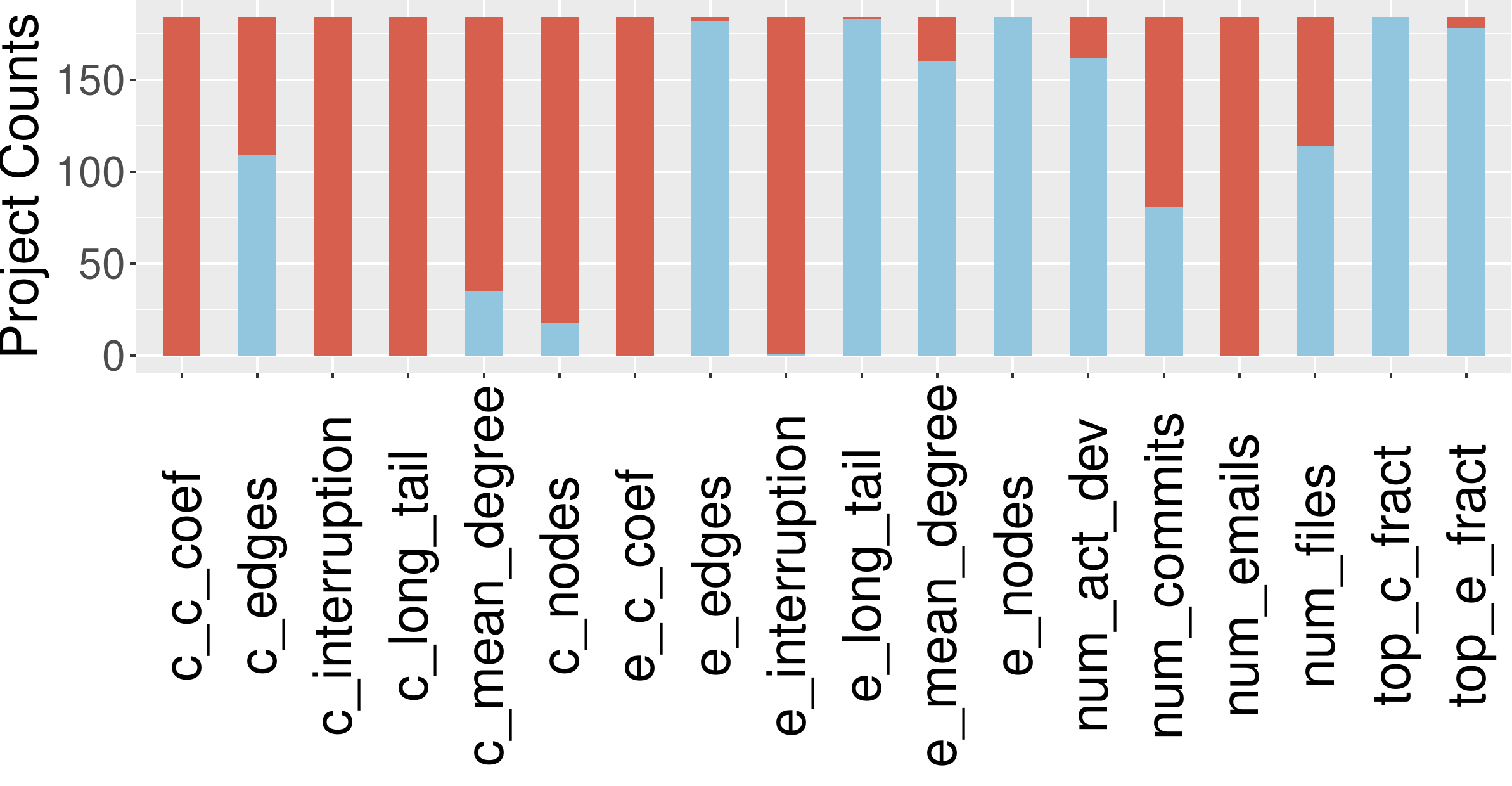}
    \caption{The overall-level coefficients of all variables of interest (blue is positive, while red is negative to graduation). It shows that some variables have same effect on almost all projects, while others do not}
    \label{fig:lime_result3}
\end{figure}

In Figure~\textcolor{red}{\ref{fig:lime_result3}}, we show the count of aggregate signs of feature coefficients across all projects.
There, \emph{blue} indicates a positive effect and \emph{red} a negative one. 
Visually, if most of the bar is a single color, then that variable has a consistent effect direction, i.e., same sign coefficient, among all projects. 
%E.g., the skewness of emails activities ($e\_skew$) is positively associated with graduation, while the skewness of commit activities ($c\_skew$) has a negative effect, across almost all projects. 
%Combine of the two results, it suggests that dense communications in the beginning of the incubation and dense commits activities around the graduation increase the graduation forecast of ASFI projects. 
Overall, perhaps surprisingly, we find that for almost all projects, the number of active developers \texttt{num\_act\_devs} is positive, the number of nodes in the technical networks \texttt{c\_nodes} has a negative effect on graduation for minor projects while the number of nodes in the social networks \texttt{e\_nodes} is positively associated with graduation.
This is consistent with prior research findings that communication is more determinant of success and onboarding than coding activities~\cite{casalnuovo2015developer}.
Other variable appear to have inconsistent effect across projects, e.g., \texttt{num\_files} and \texttt{c\_mean\_degree}.

\begin{figure}[tb]
\centering
\includegraphics[width=0.9\linewidth]{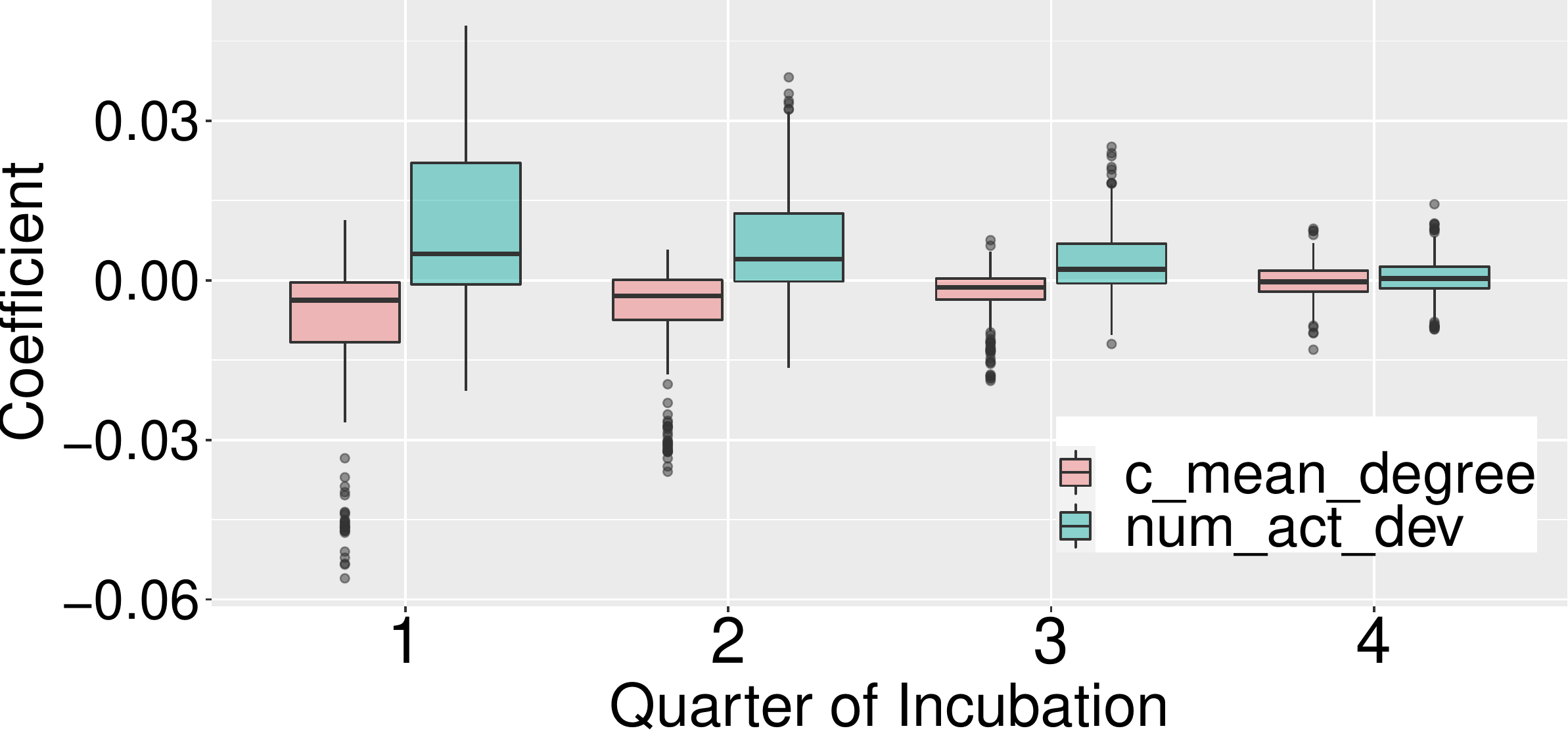}
\caption{The overall-level coefficient of two selected variables: number of active developers (in red) and mean degree in technical network (in green) in different incubating quarters}
\label{q_cmd} 
\end{figure}

Since LIME fits model coefficients for all months, we can also examine the dynamics of feature coefficients.
Figure~\textcolor{red}{\ref{q_cmd}} shows that when broken down into 4 intervals, the effect of $num\_act\_devs$ becomes less positive, and less important, over time, perhaps due to the project becoming more stable. Contrariwise, the negative effect of the mean degree in technical networks ($c\_mean\_degree$), diminishes in latter development, arguably, again, because of increased project stability over time.

\begin{figure}[t]
    \centering
    \includegraphics[width=\linewidth]{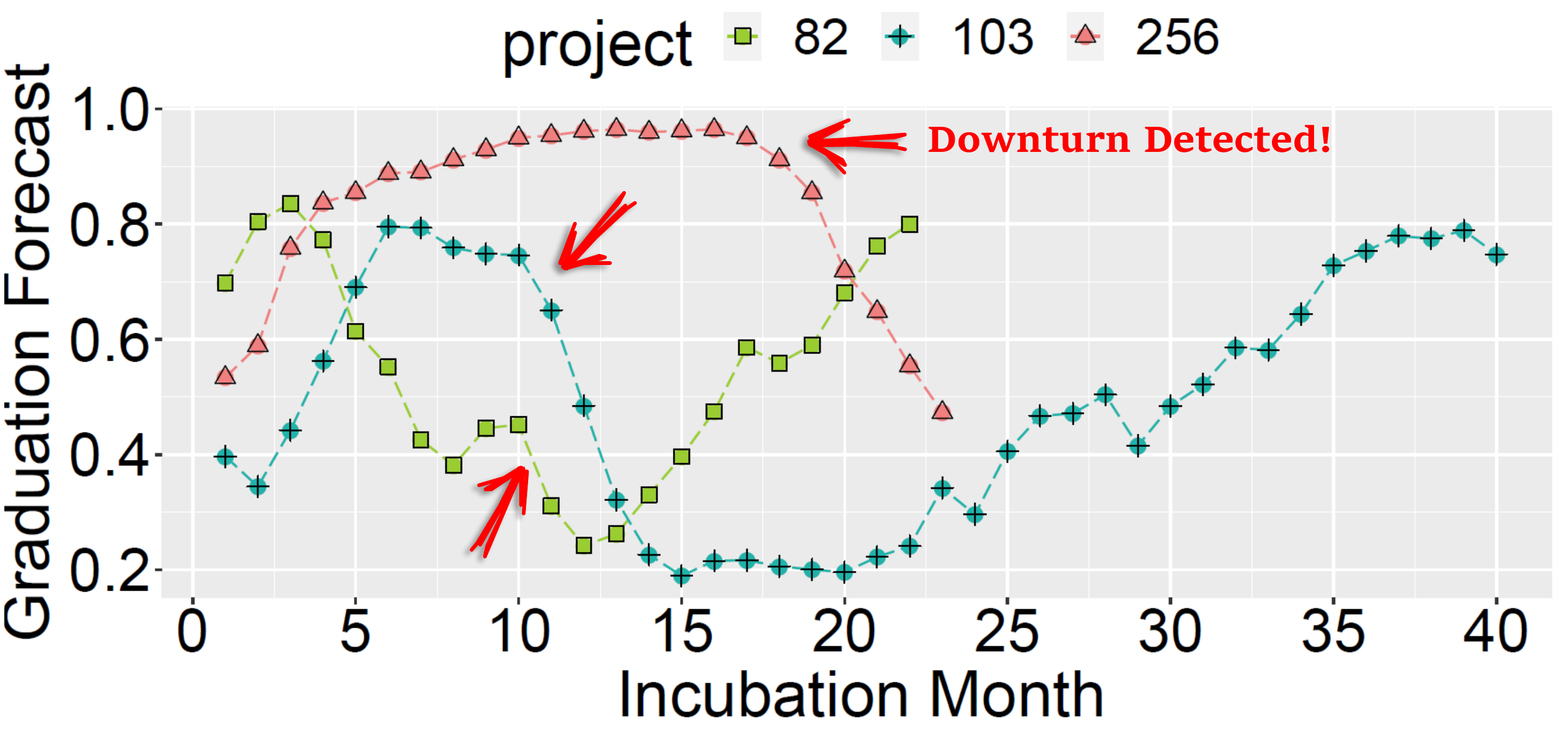}
    \caption{The graduation forecast of the marginal projects. \textit{Commonsrdf} (ID: 82, in green) and \textit{Etch} (ID: 103, in blue) are graduated projects that almost failed while retired project \textit{Ariatosca} (ID: 256, in red) almost succeeded}
    \label{case_study}
\end{figure}

%However, other variables seem to play a very different role at the overall-level, e.g, the mean degree in technical networks , 

\vspace{0.25cm}
\fbox{%
  \centering
  \parbox{0.92\linewidth}{
    \textbf{Answer to $\textbf{RQ}_{\textcolor{red}{2}}$:} 
    Effective models of project sustainability can be built from tens of socio-technical and project features. 
    Stable and interpretable models can be derived yielding feature coefficients at both project-specific and overall levels. Notably, overall, projects with \uline{fewer but more centralized committers} and \uline{those with more but distributed communicators} are more likely to become self-sustainable in the ASF incubator.
}%
}
\vspace{0.1cm}
%\vf
% \textit{The following should be a Case Study for RQ2}
\subsection{Case Study: Change of Fate}
To understand in depth why trajectories may change,  with the help of our interpretable model we identified three qualitatively different example projects, showing up or down turning points in their sustainability trajectories, as shown in Figure~\textcolor{red}{\ref{case_study}}.

\textcolor{red}{\S1}
In project \textit{Commonsrdf} (ID: 82), the graduation forecast starts high but experiences a downturn in the first half of incubation, and then it rises again. 
Our model identifies the lack of both email and commit activities are associated with the downturn.
There is also an overall decreasing trend in the number of active developers (from 27 to 11, then to 4 eventually). 
To investigate why their promising trajectory changed, and then changed again, we looked through their discussions on the mailing list. 

In the month with the lowest graduation forecast, the project released a major version of their software, following which the committers showed less activity. 
We also noticed that the ASF incubator Project Management Committee (PMC) routinely sent an email to the project asking for the monthly report~\cite{ASFI_Report}, but such reports were not requested for more than 2 months after the release. 
Lastly, we saw email arguments on the technical direction for the project's future development.
Unsurprisingly the project was going to fail if no one observes such situation and takes action to intervene. 
However, an email from one of the major contributors, \texttt{$Dev_1$}, appears to have initiated the turning point.
Below we quote it and the ensuing discussion.

\texttt{$Dev_1$} ``\textit{Folks, I've seen very little traffic for the last few months$\dots$ I am concerned that there is perhaps no longer a viable community around this podling}$\dots$'', and seriously asked ``\textit{Do people still think this project has/can build the momentum to move forwards towards graduation?}''

\texttt{$Dev_2$} ``$\dots$\textit{we lost one of the main pillars of this projects} $\dots$ \textit{So our mentors are right. We're in a situation where the project has no momentum at all, and honestly I have no idea what's best to do}$\dots$''

\texttt{$Dev_3$} ``$\dots$\textit{there was a small group of 5 core committers to begin with. As of right now, the number is $22$. We've actually done pretty well}$\dots$''

Then, \texttt{$Dev_4$} responded with an email titled `Values and Terms' and suggested a technical directions, with detailed reasoning: ``$\dots$\textit{if you actually tried to use this (algorithm) would (i) hurt speed, and (ii) hurt the perception of speed}$\dots$'' and clearly states that ``\textit{I'd be inclined to go another step further and add a generic parameter}$\dots$''

After this discussion, the community became more engaged and increased some activities, and the community felt more confident about the upcoming routine report. 
Eventually, the project graduated  in the end (our forecast went up to 80\%).
\vspace{0.05cm}

\textcolor{red}{\S2}
In project \textit{Etch} (ID: 103) there was a major depletion of senior developers after a milestone was reached in the middle of incubation. 
Then commits stopped for almost one year. 
Our graduation forecast reflects that: it dropped from 79\% to only 19\%.

After a long time being inactive, one developer sent a broadcast email titled `Future of Etch'. 
Many developers participated in the discussion thread, with seeming agreement that their project is either to be retired or changes are needed. 
The project mentor brought up the lack of diversity as a possible cause for stagnation, since all developers came from the same company. 
Some developers concurred, and they feel ``\textit{...continual pressure to wrangle new committers...}'', and consider that the ASFI values ``\textit{...extroverted tendencies of the committers rather than the merits of technology...}''. 

Eventually, the contributors reached an agreement that the project technology is and will be valuable in the future.
Among them, one developer stated that they ``\textit{...do not want to see it retired...}''
The developers then made a list of future objectives, and worked to make the community thriving again.

\textcolor{red}{\S3}
Project \textit{Ariatosca} almost succeeded, but eventually failed (the graduation forecast dropped from 96\% to 47\%). 
We find that the major reason is that all senior contributors left the project due to their busy day work. 
At the very end of the project, there were new(er) developers who wanted to contribute to the project. 
However, since newcomers could only contribute by creating Pull Requests (PRs), and PRs required a senior committer to accept and merge the code changes, they could not submit their code.
Attempts at setting meetings with the original developers failed due to busy schedules, and the project was eventually retired.

%\vspace{0.1cm}
%\fbox{%
%  \centering
%  \parbox{0.92\linewidth}{
%    \textbf{Answer to RQ4:} 
%    The proposed interpretable model can capture the downturn, and thereby providing just-in-time advice. By studying the marginals, we find the developer who openly speaks out the current situation contributes to keeping their projects sustainable. The technical regulation of ASFI make projects with no active developers have no alternative to gain community momentum.
%}
%}
\begin{comment}
\begin{figure}[tb]
    \centering
    \includegraphics[width=\linewidth]{figs/jagged_curve.pdf}
    \caption{The project \textit{Pulsar} shows a natural, jagged graduation forecast trajectory, bringing up the question: what is a large enough downturn calling for an intervention?}
    \label{jagged}
\end{figure}

\end{comment}

\subsection{$\textbf{RQ}_{\textcolor{red}{3}}$: Actionable Recommendation}
We start with a caveat. 
Precise actionable models require interventions and randomized experiments. 
Our models are not based on such experiments, and any actionability we derive from them must, therefore, be less powerful than those.
At best, our experiments can be considered \emph{natural experiments}, a subclass of quasi-experiments where the class assignment is not controlled by the experimenter. 
Thus, any interventions we suggest here must be validated experimentally in order to avoid large uncertainties in outcomes.
With that caveat in mind, we sought to answer, to the best of our experimental methods, the following intervention question: 
\underline{``What action should a project take and when?''}, in order to increase its graduation forecast in our model.

Here we propose a pragmatic, laissez-faire-unless-needed prospective strategy: to continuously monitor the graduation forecast for significant downturns, and if detected, suggest interventions that may improve the forecast.
We deconstruct the intervention question above into two parts: 1) What is a significant downturn? and 2) how to interpret the variables and coefficients in our fitted model into actions.
For the following, recall that our interpretable sustainability model gives a graduation forecast from the historical project trace data and the socio-technical project structure, available until that time.
It also yields the coefficient of each significant model feature along with its direction for every month.

\begin{figure}[t]
    \centering
    \includegraphics[width=\linewidth]{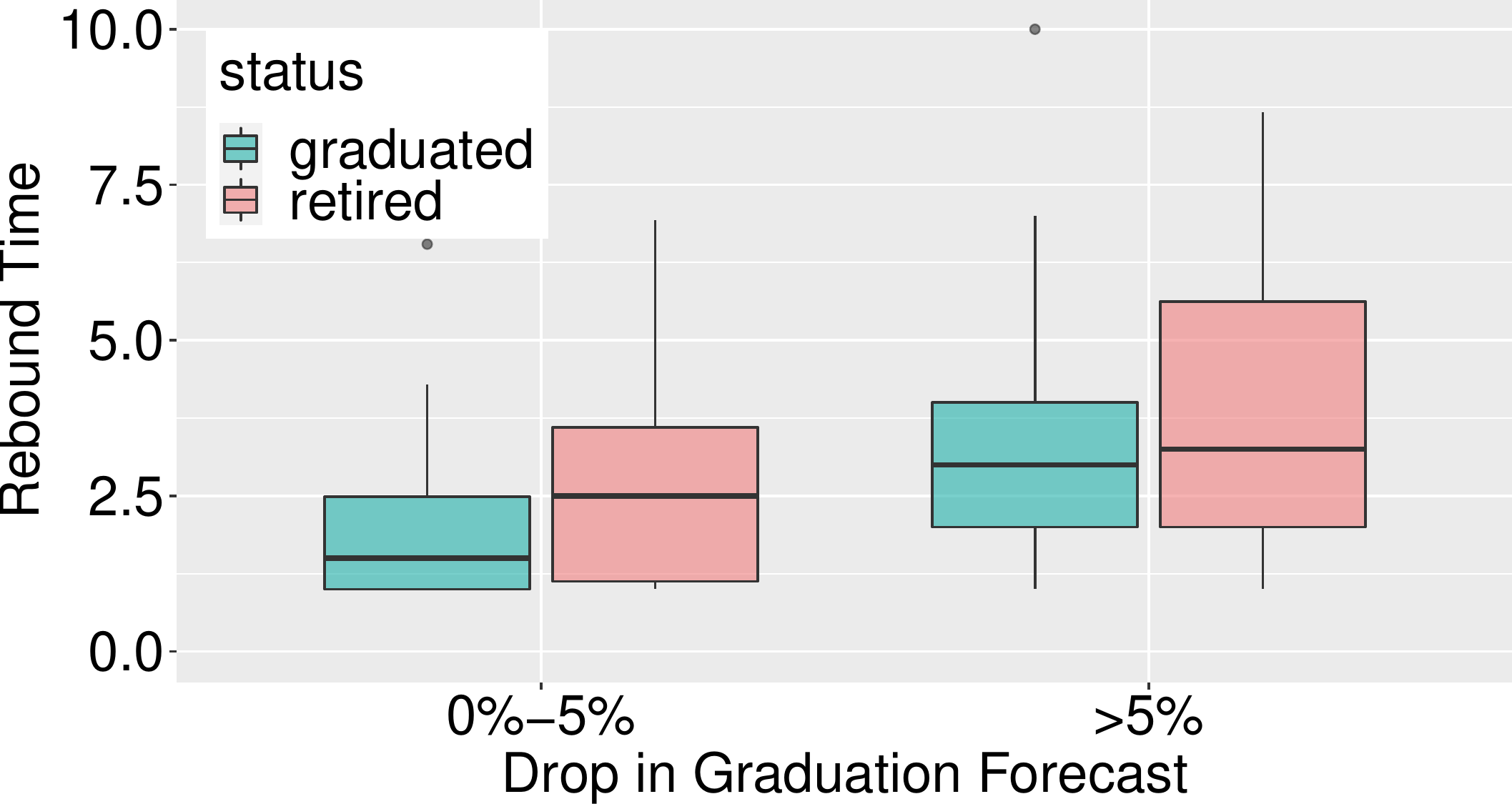}
    \caption{Bounce-up after downturn in graduation forecasts for graduated (green) and retired projects (red)}
    \label{rebound}
\end{figure}

{\underline{\textcolor{red}{\S 1} Identifying significant downturns.}} We want to identify  downturns that dominate any naturally occurring noise or jitter in the forecast. 
%e.g., see Figure~\textcolor{red}{\ref{jagged}}. 
We looked over all projects for how long it takes for a forecast to bounce-up from any downturn. 
 Figure~\textcolor{red}{\ref{rebound}} shows that the median bounce-up time is about 2.5, and, respectively, 3.5 months for drops of 0\% - 5\%, and, respectively,  $>5\%$ in the graduation forecast. 
We also noted that graduated projects seem to bounce-up faster than retired projects.
So, we use the median of the latter as the baseline, and define a drop in the forecast of greater than 5\% over a period of one or two months to indicate a significant downturn event.
This ad hoc approach, while an approximation, is a natural signal processing way to account for inherent uncertainty in the signal.

{\underline{\textcolor{red}{\S 2} Actions that improve the forecast.}} 
%We need to know how to associate actions to our model results.
Our model yields real- numbered coefficients for each significant feature in each month,  which we aggregate for stability over all months, and obtain the medians as in Fig. 4.
Increasing the values of features with positive coefficients and/or decreasing the values of the negative coefficient features results in an increase of the graduation forecast.
%As discussed at the beginning of this section, without randomized control experiments we cannot be precise about the size of the effects.
Thus, once a month or two with a significant downturn is detected,  the project developers can look at the most positive and most negative medianed significant features from the fitted model and consider increasing, respectively decreasing, them.
This is an approximation and is valid to the extent that the median is representative of the values over all months, which is more the case in the earlier months than the latter ones, see Figure~\textcolor{red}{\ref{q_cmd}}. The earlier months are the ones we care more about in practice, as we care about being most helpful to nascent projects.

Some of the socio-technical and project features may be difficult to interpret in practice.
To aid with this step, we suggest a project should compile an \emph{action table} to summarize possible actions that may positively (or negatively in the reverse way) change the value of model features.
(Multiple actions may affect the same feature.)
In Table~\textcolor{red}{\ref{actionable_table}} we provide one such action table: a  mapping between all of our model features and a non-exhaustive set of actions that we identified as likely to move each variable in the positive direction.
(The negative actions are not shown, but are complements of those.)
E.g., the \texttt{e\_c\_coef}, which counts the number of triangles in the social network, can be increased by increasing emails to everyone and not just the prominent developers or thread starters; conversely, communicating hierarchically in a tree-like fashion would decrease e\_c\_coef as it will eliminate triangles.

\begin{table}[tb]\centering
\caption{Positive Actions for Each Feature}\label{actionable_table}
\small
\begin{tabular}{ll}\toprule
 & \textbf{Positive Action (+)} \\
\hline
%\multicolumn{3}{c}{\textbf{Project Metrics}}\\
num\_act\_dev 
& Contribute frequently; Advertise, Recruit.\\
num\_emails
& Reach out; Ask questions; Encourage communication.\\
num\_commits 
& Commit frequently; commit smaller; use CI. \\
num\_files 
& Split files; refactor code; encourage modularity. \\
c\_interruption 
& Go on vacation often; contribute in bursts.\\
e\_interruption 
& Email seldom; discourage discussion.\\
top\_e\_fract 
& Encourage core emailers to respond more. \\
top\_c\_fract  
& Core contributors commit exclusively. \\
\hline
%\multicolumn{3}{c}{\textbf{Technical Networks (TNs)}} \\
c\_nodes 
& Establish technical mentorship; encourage commits. \\
c\_edges 
& Commit to same files as others; document code well.\\
c\_c\_coef 
& Encourage collaborations, pair programming. \\
c\_mean\_degree 
& Encourage commits by minor contributors. \\
c\_long\_tail 
& Mentor collaborations with newcomers. \\
\hline
%\multicolumn{3}{c}{\textbf{Social Networks (SNs)}} \\
e\_nodes 
& Mentor low communicators.\\
e\_edges 
& Reply to questions; ask questions.\\
e\_c\_coef 
& Encourage non-hierarchical communications. \\
e\_mean\_degree 
& Communicate with minor emailers. \\
e\_long\_tail 
& Foster communication-heavy culture.  \\
\bottomrule
\end{tabular}
\end{table}

\begin{figure}[t]
    \centering
    \includegraphics[width=\linewidth]{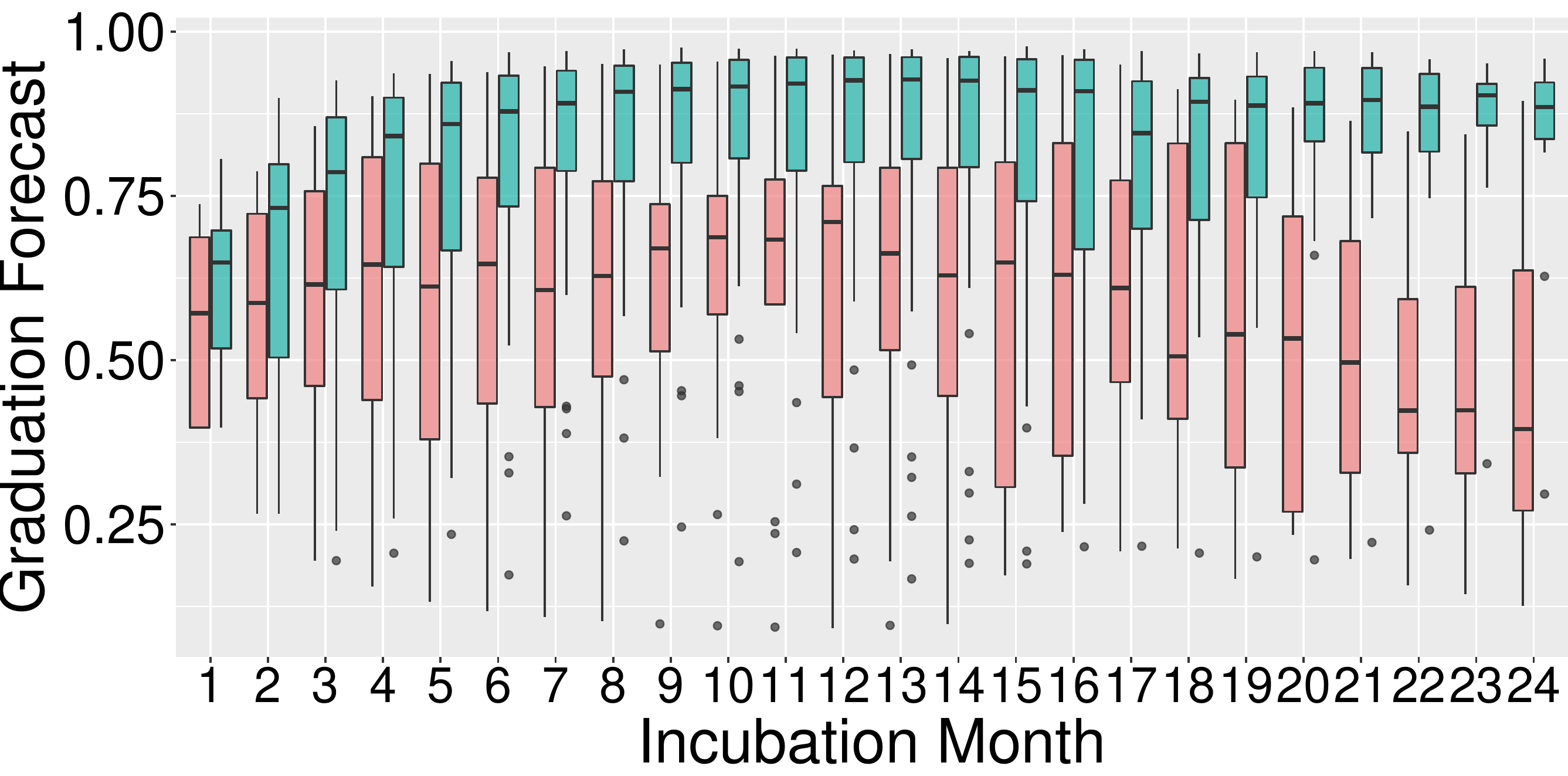}
    \caption{Graduation forecasts for all graduated (green) and retired projects (red) over the first 24 incubation months}
    \label{sr}
\end{figure}

{\underline{Ecosystem and project-specific fine-tuning.}} Our strategy can be further fine-tuned in several ways.
First, there are common patterns in the graduation forecasts over all projects.
Figure~\textcolor{red}{\ref{sr}} shows that for graduated projects (in green) there is an apparent upward trend in the forecast in the first $6$ months, suggesting that the early-stage development deserves more attention from project managers. 
From month $6$ to month $12$ we do not observe a significant change, and the decreasing variance also tends to support such an argument. 
However, we find increasing variance in months $12$ through $18$. One possible reason is that many graduated projects achieve their milestone in that period, and slow further commits and discussions, thus lowering the graduation forecast. 
%
%In summary, many graduated projects already show their graduating trend within their first $6$ months, the first and significant downturn starts $12$ months after the incubation begins. 
These considerations can be taken into account during forecast monitoring, with more frequent monitoring chosen or increased attention paid at times around milestones and releases.
The ASF committees can also ask for more frequent project reports, during the first $6$ months and $12$ months of incubation, as project reports were seen to be an incentive to productivity in our case study. 

We recognize that some socio-technical elements are more difficult to change compared to others.
This is, in general, project-specific, in that some projects can easier modify some features than can other projects.
E.g., if the interpretable model suggests increasing \texttt{c\_edges} and decreasing \texttt{c\_interruptions} this may be easier done in smaller projects than larger ones due to the difficulty of influencing many people at once.  
Thus these action tables should ideally be project-specific, designed and updated as the project evolves.

Lastly, a word of caution. In terms of expectations that including more features to intervene on, may lead to faster bounce-up, we note that empirical evidence shows socio-technical features have ways of getting correlated pairwise over time in the same system.
%E.g., if projects can recruit more new committers, it might improve the situation indirectly.
Thus, even if not correlated, the effect of increasing multiple features simultaneously may not be additive.

\vspace{0.25cm}
\fbox{%
  \centering
  \parbox{0.92\linewidth}{
    \textbf{Answer to $\textbf{RQ}_{\textcolor{red}{3}}$}: 
    Our strategy can be used as a monitoring tool that feeds suggestions into developers' decision process.
    Project-specific features can be selected from the suggestions for intervention when experiencing downturns.
    The algorithm can be made bespoke by introducing more frequent monitoring around releases/milestones.
}
}
\vspace{0.25cm}

\subsubsection{Actionable Strategy Example}
Here we apply our strategy on a project from Figure~\textcolor{red}{\ref{case_study}}: \textit{Commonsrdf}, and show specific recommendations following a detected downturn.

While monitoring project \textit{Commonsrdf} \footnote{\url{http://mail-archives.apache.org/mod\_mbox/commonsrdf-dev/}} 
we would have observed  a significant downturn at months 4 and 5 (> 5\% drop), see Figure~\textcolor{red}{\ref{case_study}}. 
For that project, our interpretable model yields as the 3 features with the highest positive medians over all months: \texttt{top\_c\_fract}, \texttt{top\_e\_fract}, and \texttt{e\_nodes}; the three with the most negative median are:  
\texttt{c\_c\_coef}, \texttt{c\_interruption}, and \texttt{c\_nodes}.
Consulting Table~\textcolor{red}{\ref{actionable_table}}, it calls for the project to increase the commit and email contributions by core developers and encourage more committers to communicate.
It also calls for the project to decrease collaborations, decrease commit interruptions, and decrease the number of developers that commit. 
A continuous integration may also be recommended to this project, to keep people on track with smaller, more frequent commits.

What actually happened in the project is that that initial period of downturn was missed; as we saw in our case study for this project, the project manager sent an email titled 'Anybody there?' in month 8, when productivity was already significantly reduced.
Using our strategy, the downturn could have been identified and possibly avoided 3 months sooner.

\section{Threats to Validity and Conclusion}

\uline{Threats.} First, our commit and email data is from only hundreds of projects ASF incubator projects. 
%As an inherent dis-balance exists between graduated and retired projects, there may be bias in our forecasting algorithm.
Thus, generalizing the implications beyond ASF, or even beyond the ASF Incubator projects carries potential risks. Expanding the dataset beyond ASF, e.g., with additional projects from other open-sourced incubator projects can lower this risk. 
Second, we do not consider communications other than through the ASF mailing lists. However, ASF's policies and regulations insist on the use of mailing lists, which lowers this risk. 
Lastly, interpreting deep learning models is still an art, and
LIME models are approximations.
They may be particularly sensitive to correlated features. We lower such risk by eliminating correlated variables before training.
Taking the actionable suggestions given in this paper may result in changes of more than one variable, e.g., increase the active developers may also increase the number of commits.

\uline{Conclusion.} 
Understanding why many nascent projects have failed may help others improve their individual practice, organizational management, and institutional structure. 
Here we showed that quantitative network science approaches combined with state-of-the-art AI methods can effectively model ASF incubator project sustainability, from a novel longitudinal dataset of socio-technical contributions in ASFI projects, more narrow in scope than general OSS projects but with extrinsic graduation/sustainability labels. 
We also demonstrated the combined power of mixed methods: through case studies, we identified specific reasons for success and failure of projects, complementing our models. 
Finally, we developed a strategy for translational use of the models in practice. Our methods make it straight forward to track a project's trajectory as it progresses toward sustainability, and even offer advice for correcting trajectories upwards.
Future work is needed to offer validation of this or similar strategies experimentally.

\section*{Acknowledgements}
We are grateful to the National Science Foundation for funding this project, under grant \#2020751. We greatly thank the FSE 2021 reviewers for their constructive comments which helped improve this paper.

\bibliography{refs}
\bibliographystyle{acm}

\end{document}